\documentclass[12pt]{emulateapj}

\def\etl{et al.}

\def\niilam{[\ion{N}{2}]~$\lambda6584$}
\def\niidoublet{[\ion{N}{2}]~$\lambda\lambda6548,6583$}
\def\siidoublet{[\ion{S}{2}]~$\lambda\lambda6717,6731$}

\def\oiii{[\ion{O}{3}]}
\def\oiiilam{[\ion{O}{3}]~$\lambda5007$}
\def\oiiidoublet{[\ion{O}{3}]~$\lambda\lambda4959,5007$}

\newcommand{\mr}{\ensuremath{M_{r}}}
\newcommand{\omegam}{\ensuremath{\Omega_{\rm m}}}
\newcommand{\omegal}{\ensuremath{\Omega_{\Lambda}}}
\newcommand{\hubble}{\ensuremath{h_{100}}}
\newcommand{\hh}{\ensuremath{H_{0}}}
\newcommand{\kms}{\ensuremath{{\rm km~s}^{-1}}}
\newcommand{\mpc}{\ensuremath{{\rm Mpc}^{-1}}}

\newcommand{\Ae}{\ensuremath{{\alpha}{\rm -enhancement}}}
\newcommand{\Hd}{\ensuremath{{\rm H}\delta}}
\newcommand{\HdA}{\ensuremath{{\rm H}\delta_{\rm A}}}
\newcommand{\HdF}{\ensuremath{{\rm H}\delta_{\rm F}}}

\newcommand{\CAone}{\ensuremath{{\rm Ca}}4227}

\newcommand{\Hg}{\ensuremath{{\rm H}\gamma}}
\newcommand{\HgA}{\ensuremath{{\rm H}\gamma_{\rm A}}}
\newcommand{\HgF}{\ensuremath{{\rm H}\gamma_{\rm F}}}

\newcommand{\Ctwo}{\ensuremath{{\rm C}_2}4668}

\newcommand{\Hb}{\ensuremath{{\rm H}\beta}}

\newcommand{\Mgb}{\ensuremath{{\rm Mg}\, b}}
\newcommand{\Fefour}{\ensuremath{{\rm Fe}}5270}
\newcommand{\Fefive}{\ensuremath{{\rm Fe}}5335}

\newcommand{\Ha}{\ensuremath{{\rm H}\alpha}}

\newcommand{\Fe}{\ensuremath{\langle {\rm Fe}\rangle}}
\newcommand{\MgFep}{\ensuremath{[{\rm MgFe}]^{\prime}}}
\newcommand{\MgFe}{\ensuremath{[{\rm MgFe}]}}

\newcommand{\aFe}{\ensuremath{\alpha/{\rm Fe}}}

\newcommand{\osigma}{\ensuremath{1\sigma}}

\newcommand{\latin}[1]{{#1}}

\newcommand{\eg}{\latin{e.g.}}

\newcommand{\Sersic}{S\'ersic}

\usepackage{natbib}

\begin{document}
\shorttitle{Stellar Populations of Ellipticals}
\shortauthors{Zhu \etl}
\title {Stellar Populations of Elliptical Galaxies in the Local Universe}

\author{
 Guangtun Zhu\altaffilmark{1},
 Michael R. Blanton\altaffilmark{1}, and
 John Moustakas\altaffilmark{2}
} 
\altaffiltext{1}{
Center for Cosmology and Particle Physics, Department of Physics, New York University,
4 Washington Place, New York, NY 10003, gz323@nyu.edu, michael.blanton@nyu.edu}
\altaffiltext{2}{Center for Astrophysics and Space Sciences, University of California, 
San Diego 9500 GilmanDrive La Jolla, California, 92093-0424, jmoustakas@ucsd.edu}


\begin{abstract}
We study the stellar populations of $1,923$ elliptical galaxies at
$z<0.05$ selected from the Sloan Digital Sky Survey as a function of
velocity dispersion, $\sigma$, and environment.  Our sample
constitutes among the largest high-fidelity samples of elliptical
galaxies with uniform imaging and optical spectroscopy assembled to
date.  Confirming previous studies, we find that elliptical galaxies 
dominate at high luminosities ($\gtrsim L^{\ast}$), and that
the highest-$\sigma$ ellipticals favor high-density environments.  
We construct average, high signal-to-noise spectra in bins of $\sigma$ and
environment and find the following: (1) lower-$\sigma$ galaxies have a
bluer optical continuum and stronger (but still weak) emission
lines; (2) at fixed $\sigma$, field ellipticals have a slightly
bluer stellar continuum, especially at wavelengths $\lesssim 4000$
\AA, and have stronger (but still weak) emission lines compared to
their group counterparts, although this environmental dependence is
strongest for low-$\sigma$ ellipticals and the highest-$\sigma$
ellipticals are much less affected.  Based on Lick indices
measured from both the individual and average spectra, we
find that: (1) at a given $\sigma$, elliptical galaxies in groups
have systematically weaker Balmer absorption than their field
counterparts, although this environmental dependence is most
pronounced at low $\sigma$; (2) there is no clear environmental
dependence of \Fe, while the $\alpha$-element absorption indices such
as \Mgb~are only slightly stronger in galaxies belonging to rich
groups.  An analysis based on simple stellar populations (SSPs)
reveals that more massive elliptical galaxies are older, more
metal-rich and more strongly $\alpha-$enhanced. We also find
that: (1) the SSP-equivalent ages of galaxies in rich groups are,
on average, $\sim 1$~Gyr older than in the field, although
once again this effect is strongest at low $\sigma$;
(2) galaxies in rich groups have slightly lower [Fe/H]
and are marginally more strongly $\alpha-$enhanced; and (3) there is
no significant environmental dependence of total metallicity, [Z/H].
Our results are generally consistent with stronger low-level
recent star formation in field ellipticals at low $\sigma$, 
similar to recent results based on ultraviolet and infrared observations.  
We conclude with a brief discussion of our results in the context of 
recent theoretical models of elliptical galaxy formation.
\end{abstract}

\keywords{galaxies: fundamental parameters (classification, colors, luminosities, masses, radii, etc.) ---
  galaxies: elliptical and lenticular, cD  ---
  galaxies: formation -- 
  galaxies: evolution --
  galaxies: stellar content}

\section {Introduction}

The formation and evolution of elliptical galaxies remains one of the
most challenging open problems in the general theory of galaxy
formation and evolution.  In the current standard $\Lambda$CDM
cosmological model \citep{komatsu10a}, structure grows hierarchically
\citep{white78b} and merging is an unavoidable process in galaxy
formation.  It has long been proposed that spiral galaxies may
eventually merge and form elliptical galaxies \citep{toomre77}.
Recent improvements in both observations and numerical simulations
have yielded remarkable support for the merging picture.
Deep photometry \citep{ferrarese94, kormendy94, lauer95, kormendy99,
lauer05, ferrarese06b, lauer07, cote07, kormendy09} with the {\it
Hubble Space Telescope} (HST) and integral-field spectroscopy
\citep{bacon01, emsellem07, cappellari07} have shown that lower-mass
elliptical galaxies with $M < M^{\ast}$ in general have cuspy
(extra-light) surface brightness profiles in their centers and are
kinematically supported by relatively fast rotation, while more
massive elliptical galaxies with $M > M^{\ast}$ have ``core-like''
central surface brightness profiles (i.e., missing light), and are
usually slow rotators.  Recent numerical studies \citep{mihos94c,
cox06b, naab09a, hopkins09a} have shown that gas-rich mergers
between disk galaxies (wet mergers) can produce a cuspy central
surface brightness profile and the fast rotational kinematic signature
of low-mass elliptical galaxies, while subsequent gas-poor mergers
between less massive elliptical galaxies (dry mergers) then will form
more massive elliptical galaxies \citep{naab06a, hopkins09b}.

The mass and luminosity functions of massive red galaxies from deep
high redshift surveys at $z\sim1$ also suggest that dry mergers could
have played an important role in elliptical galaxy formation since
redshift one \citep{bell04a, faber07}.  However, whether or not dry
mergers are important remains controversial.  Other studies show that
massive red galaxies may have not undergone many dry mergers since
redshift unity \citep{cimatti06, brown07, cool08}.  Merger rate
studies also draw various conclusions about the significance of dry
mergers \citep{bell04b, bell06b, bell06a, masjedi06, masjedi08, robaina10a}.
Meanwhile, studies of early-type
galaxies\footnote{In this paper, we refer to elliptical and lenticular
galaxies (E/S0) as early-type galaxies, but we only include
elliptical galaxies (E) in our sample.} at very high redshift show
that a significant fraction of massive evolved spheroidal stellar
systems are already in place at very high redshift ($z\gtrsim2$).
Most of them appear to be very compact \citep{daddi05, longhetti07,
  toft07, trujillo07, vandokkum08, cimatti08, saracco09} and it is
possible that through minor mergers or gradual accretion they can
evolve to elliptical galaxies at the present day \citep{naab09b,
  vandokkum10}.

An alternative way to place strong constraints on elliptical galaxy
formation theory is through the detailed study of the local universe.
Such work has been extensively undertaken for decades.  Besides the
deep photometric surface brightness profile and integral-field
spectroscopic kinematic studies cited above, a non-exhaustive list
includes: various scaling relations (\eg, the Faber-Jackson relation,
the fundamental plane, etc.) \citep{faber76, dressler87,
  djorgovski87}; the color-magnitude diagram (\citealt{faber73,
  bower92, blanton03d, hogg04, baldry04, balogh04}); and
absorption-line indices (\citealt{peletier89, worthey92, jorgensen99,
  trager00b, thomas05}, hereafter T05; among many others).  These
studies show that elliptical galaxies have remarkably uniform
properties and that their stellar mass content is dominated by old
stellar populations.

The study of environmental effects can also shed light on the theory
of elliptical galaxy formation and evolution.  It has been
known for some time that the most massive and luminous galaxies
favor high-density regions \citep{dressler80a, binggeli88a,
hogg03, hogg04, kauffmann04, croton05, blanton05b, blanton09},
lending support to the hierarchical model \citep[\eg,][]{mo96,
lemson99, berlind05}.  Numerous studies have also
investigated the environmental dependence of the  scaling
relations.
For example, the fundamental plane for early-type galaxies
seems to show a small dependence on environment
\citep[\eg,][]{bernardi06, labarbera10c}.  The color-magnitude
relations in different environments also exhibit a weak, though
statistically significant difference \citep[\eg,][]{bernardi03b,
hogg04, skibba09b, blanton09}.  A number of studies show that the index-velocity
dispersion ($\sigma$) relations such as Mg-$\sigma$ and Fe-$\sigma$
show no or a weak dependence on environment (\eg,
\citealt{jorgensen97, trager00b, sanchez03}, among others).  However,
interpretation of these weak dependencies is complicated.  T05, using
a sample of 124 E/S0 galaxies and recent simple stellar
populations (SSP) models \citep[][hereafter TMB]{thomas03}, find
that massive early-type galaxies in low-density environments are on
average $\sim 2$ Gyr younger and slightly more metal-rich than their
counterparts in high-density environments.  \citet[]{sanchez06b},
using a sample of 94 E/S0 galaxies reported similar results.

One surprising recent finding is that many elliptical galaxies seem to
have a non-negligible fraction of young stars.  Absorption line
studies and fundamental plane studies favor a ``frosting'' model in
which early-type galaxies consist of an old base population with a
small amount of younger stars (\eg, \citealt{trager00a, gebhardt03,
schiavon07}, S07 hereafter). 
Results from the {\it Galaxy Evolution Explorer}
(GALEX; \citealt{martin05}), the {\it Spitzer Space Telescope} \citep[][]{werner04a}
and the HST have shown that a significant
fraction of early-type galaxies exhibit strong ultraviolet (UV)
excess, polycyclic aromatic hydrocarbon (PAH) emission and infrared
(IR) excess, implying possible low-level recent star formation
\citep{yi05, rich05, kaviraj07b, schawinski07, temi09, young09,
  salim10}.  Recent star formation is also consistent with observed
cold gas in many ellipticals (e.g., \citealt{faber76b, knapp85a,
  vangorkom97a}).  \citet{schawinski07} find that the fraction of
near-UV (NUV) bright early-type galaxies is $\sim 25\%$ higher in
low-density environments, possibly due to stronger low-level recent
star formation.  \citet{kaviraj09} suggest that minor mergers can
account for the inferred amount of star formation.  Using
spatially resolved spectroscopy, \citet{shapiro10} 
\citep[see also][]{kuntschner10a} find that star
formation in early-type galaxies happens exclusively in fast-rotating
systems and occurs in two distinct modes: the first with widespread
young stellar populations associated with a high molecular gas
content, and the second with disk or ring morphology (see also
\citealt{young02a, young05a, young08a}).  They suggest the first may
be due to minor mergers, and the second due to rejuvenation in
previously quiescent stellar systems.

The Sloan Digital Sky Survey \citep[SDSS,][]{york00} has provided the
largest set of local galaxies with uniform imaging and spectroscopy
and offers a great opportunity to study the nearby elliptical galaxies
in a systematic and homogeneous way.  Previous work using the SDSS
relied on the pipeline outputs in the SDSS to select early-type
galaxies \citep[\eg,][]{bernardi03a, eisenstein03}.  This method does
not only select elliptical galaxies, but also lenticular galaxies and
early-type spiral galaxies, which may have different properties and
formation pathways.  Because the SDSS provides high-quality
imaging for nearby galaxies, we reanalyze these images, use their
detailed surface brightness profiles to preselect bulge-dominated
galaxies, and visually examine the bulge-dominate galaxies to select a
high-fidelity clean sample of $1,923$ elliptical galaxies.  Our
sample constitutes among the largest high-fidelity samples of
elliptical galaxies with uniform imaging and spectroscopy assembled
to date.  Comparable samples are the visually selected
early-type galaxy samples from the Galaxy Zoo project
\citep[\eg,][]{schawinski09} and the MOrphologically Selected
Early-types in the SDSS project (MOSES, \citealt{schawinski07,
thomas10}) --- however, these samples are at larger distances than
ours ($z>0.05$), which at the spatial resolution of the SDSS images
makes the classification problem more difficult.  Using this
elliptical sample, we study the dependence on $\sigma$ and environment
of the stellar populations in elliptical galaxies.
 
The rest of this paper proceeds as follows.  In $\S$2, we describe our
parent sample and our method for selecting a high-fidelity sample of
elliptical galaxies.  In $\S$3, we present the photometric properties
of our sample and determine their local environments.  In $\S$4, we
study the average optical spectra, Lick indices, metallicity and age,
and their dependence on $\sigma$ and environment.  We discuss our
results in the context of theoretical models in $\S$5 and summarize
our principal conclusions in $\S$6.
 
We adopt a $\Lambda \mathrm{CDM}$ cosmology with $\omegam=0.3$,
$\omegal=0.7$ and $\hh=100~\hubble~\kms~\mpc$ with $\hubble=0.7$.  All
apparent magnitudes are on the native SDSS photometric
system and all absolute magnitudes are on the AB system and corrected
for Galactic extinction \citep{schlegel98} and K-corrections
\citep{blanton07}.
 
\section{Data and sample selection}
 
\subsection{Parent sample}\label{sec:parent}

The SDSS-I and SDSS-II imaged $11,663$ square degrees of the sky in
$ugriz$ and obtained optical spectra for $\sim 1.6$ million objects,
$\sim 0.7$ million of which are galaxies with $r<17.77~\mathrm{mag}$
\citep[\eg,][]{gunn98,york00,strauss02,abazajian03}.  Automated
software performs all of the data processing: astrometry
\citep{pier03}; source identification, deblending and photometry
\citep{lupton01}; photometricity determination \citep{hogg01};
calibration \citep{fukugita96,smith02}; spectroscopic target selection
\citep{eisenstein01,strauss02,richards02}; spectroscopic fiber
placement \citep{blanton03a}; and spectroscopic data reduction.  More
detailed descriptions of these pipelines can be found in
\citet{stoughton02}.  An automated pipeline called {\tt
  idlspec2d}\footnote{\tt
  http://spectro.princeton.edu/idlspec2d\_install.html} measures the
redshifts and classifies the reduced spectra.

For the purposes of computing large scale structure and galaxy
property statistics, \citet{blanton05a} have assembled a subsample of
SDSS galaxies known as the NYU Value Added Galaxy Catalog
(NYU-VAGC)\footnote{\tt http://sdss.physics.nyu.edu/vagc/}.  We select
our parent sample from the NYU-VAGC low-redshift catalog, consisting
of all galaxies in the SDSS with $z<0.05$.  We use the version of this
catalog corresponding to SDSS Data Release 6 (DR6,
\citealt{adelman06}), which contains $77,149$ galaxies.

Due to the difficulty of automatic photometric processing of big
galaxies, the SDSS catalog is missing many nearby, bright galaxies,
even though they are contained within the SDSS imaging footprint; the
incompleteness begins to become important at
$r\lesssim14.5$.  Therefore, to ensure a complete parent sample, we
include any low-redshift galaxies from the Third Reference Catalog of
Bright Galaxies (RC3; \citealt{devaucouleurs91, corwin94}) for which
we have $ugriz$ imaging from the SDSS, but which are not in the
NYU-VAGC.
Including $10,474$ galaxies from the RC3, our combined parent sample
contains $87,623$ galaxies at $z<0.05$ ({\tt LowZ}; see Table
\ref{sampletable}).

We expect all bona fide elliptical galaxies to occupy the red sequence
of the color-magnitude diagram, since even a modest amount of ongoing
star formation will result in a blue optical color.  Therefore, we
isolate galaxies on the red sequence using a luminosity-dependent
color cut:
\begin{equation}
\label{colorcut}
M_g-M_i > -0.05\times(M_r+16.0)+0.65 \mathrm{.}  
\end{equation} 
This cut is very generous in that it extends all the way to the edge
of the blue cloud, and should therefore include most or all of the
blue elliptical galaxies found by recent studies (e.g.,
\citealt{fukugita04, schawinski09}).  There are $37,026$ galaxies that
satisfy equation (\ref{colorcut}), which we define as our photometric
red-sequence parent sample ({\tt PhotoRS}).  Within this sample,
$32,726$ have spectroscopy from the SDSS, which we define as our spectroscopic
red-sequence parent sample ({\tt SpecRS}; see
Table~\ref{sampletable}).

\subsection{Elliptical sample selection}\label{sec:selection}

Traditionally, galaxies are classified into different morphological
types by visual inspection \citep{devaucouleurs59, sandage75, smail97,
  desai07, lintott08}.  Unfortunately, visual classification is
inherently subjective, and not feasible for samples consisting of tens
of thousands of galaxies, which has led to a concerted effort by many
different groups to classify galaxies using objective, quantitative
criteria \citep[e.g.,][]{conselice03a, lotz04a, scarlata07a}.
However, although quantitative morphological classification schemes
have become increasingly sophisticated, they arguably do not capture
the detailed variation in morphological signatures apparent to the
trained eye (see the recent discussion in \citealt{blanton09}, and
references therein).  Moreover, in studies of elliptical galaxies,
contamination from early-type disk galaxies (i.e., S0/Sa), can be
significant.  Indeed, objective classification methods have tremendous
difficulty distinguishing bulge-dominated disk galaxies from true
elliptical galaxies, although the two galaxy types may have
experienced very different formation pathways. 

Given these various issues, we opt for a hybrid approach, in which we
conservatively \emph{preselect} elliptical galaxies using well-defined
quantitative criteria, and then use visual inspection to remove
contaminants.  Thus we retain the objectivity of quantitative methods
and the ability to analyze a large parent sample, while still relying
on the superb ability of the eye to identify subtle but important
morphological signatures like faint spiral arms, dust lanes, bars,
etc., to help remove contaminants.  In the next three sections we
describe our elliptical galaxy selection in more detail.

\subsubsection{The {\it Ellipse} method}

We first reduce the two-dimensional (2D) images to one-dimensional
(1D) surface brightness profiles using the {\tt Ellipse} algorithm
(\citealt{jedrzejewski87}, see also \citealt{young79, kent83}).  The
{\tt Ellipse} method is widely used in the imaging analysis of
galaxies because the isophotes of galaxies are well approximated by
ellipses.

The basic idea of the {\tt Ellipse} method is to expand the intensity
around an ellipse in a Fourier series (see \citealt{bender89} for the
expansion in polar coordinates):
\begin{equation}
  I (E) = I_0 + \sum_{n=1} [A_n \cos(nE) + B_n \sin(nE)] \mathrm{,}
\end{equation}
where $I$ indicates the intensity, $E$ is the eccentric anomaly, and
$A_n$ and $B_n$ are the higher-order Fourier coefficients.  For a
given major axis with length $a$, we use $n=1$ and $n=2$ to find the
ellipse that best matches the measured isophote.  Specifically, we
first provide an initial guess for the center of the ellipse ($x_0,
y_0$), the ellipticity ($\epsilon$), the position angle ($\phi$), and
the intensity ($I_0$).  We then calculate $A_n$ and $B_n$ ($n=1,2$)
and update the parameters by calculating the expected deviation from
the true parameters (see \citealt{jedrzejewski87}), and iterate until
the maximum of $A_n$ and $B_n$ ($n=1,2$) is less than $4\%$ of the
intensity ($I_0$).  Next, we measure the third- and fourth-order
harmonics of the resulting intensity distribution using least-squares
minimization.  If an isophote is a perfect ellipse then all the
coefficients, $(A_n, B_n), n=1, \cdot\cdot\cdot, \infty$ will be
identically zero.  Non-zero coefficients indicate the amount by which
the isophote deviates from the shape of an ellipse.  In particular,
$A_4 > 0$ indicates a ``disky'' isophote, while $A_4 < 0$ corresponds
to a ``boxy'' isophote \citep[\eg,][]{bender89}.
 
We apply the {\tt Ellipse} method to each galaxy image in all five
bandpasses as a function of the major axis radius, using a one-pixel
step size ($0.396$ arcsec) along the major axis.  At each step, we
save all the parameters, including the position of the origin ($x_0,
y_0$), the major axis length ($a$), the ellipticity ($\epsilon$), the
position angle ($\phi$), the intensity ($I_0$), the harmonic
coefficients ($A_n$ and $B_n$ with $n=1,2,3,4$), and the total flux
within each ellipse.
 
\subsubsection{Bulge-disk decomposition}

Given the {\tt Ellipse} fitting results, we model the
surface brightness profile, $I_{R,j}$, in each bandpass $j=g, r, i$
simultaneously as a function of radius $R$ using a two-component
``bulge plus disk'' model:

\begin{equation}
  I_{R,j} = I_{B,j} \exp (-K_{B} R^{\frac{1}{n}}) + I_{D,j} \exp
  (-K_{D} R),
\label{eq:model}
\end{equation}

\noindent where the first term represents a \Sersic~model of index $n$
\citep[\eg,][]{sersic63, ciotti91, graham05} for the bulge component
($B$), and the second term is an exponential model for the disk
component ($D$).  The parameters $K_{B}$ and $K_{D}$ are length-scale
factors for the bulge and disk component, respectively; they are
constrained to be equal in all three bandpasses, while the amplitudes
$I_{B,j}$ and $I_{D,j}$ are allowed to vary.  Note that we have chosen
not to include the $u$- and $z$-band surface brightness profiles in
this analysis because the photometry in those bands is considerably
noisier.

\subsubsection{Preselection and visual
  inspection}\label{sec:preselect} 

Our goal in this section is to select a high-fidelity sample of
elliptical galaxies.  As discussed in \S\ref{sec:selection}, our
strategy is to apply conservative parameter cuts to our
ellipse-fitting results to preselect a sample of spheroidal galaxies,
and then to visually inspect the resulting sample to remove
contaminating disk-dominated galaxies.

First, we require the bulge-to-total ratio, $B/(B+D)$, to be larger
than $0.7$ in all of the three bandpasses.
We then reject galaxies with ellipticities larger than $0.6$, which
only rejects a handful of real elliptical galaxies (see $\S3.1$ and
Fig.~\ref{rsmakeup}).
Next, we parametrize the ``featurelessness'' of the
surface brightness profile by fitting a straight line to the
ellipticity versus radius profile, and compute $\chi^2$ assuming
uniform errors for the ellipticities.  We define a featurelessness
parameter by multiplying the $\chi^2$ by the largest ellipticity in
the profile.  By exploring training sets, we apply a generous cut in
this parameter to reject non-elliptical contaminants while
simultaneously minimizing the number of real elliptical galaxies that
are excluded.
Finally, because the instrumental dispersion of the SDSS spectrograph
is $69~\kms$, a velocity dispersion measurement smaller than $70~\kms$
is not reliable\footnote{\tt
  http://www.sdss.org/dr7/algorithms/veldisp.html}; therefore, we
only include galaxies with $\sigma>70~\kms$.  Among the $32,726$
galaxies in our {\tt SpecRS} sample (see Table~\ref{sampletable}),
$22,621$ objects have $\sigma>70~\kms$, of which $2,648$ survive our
preselection cuts.

\begin{figure}
\epsscale{1.0}
\plotone{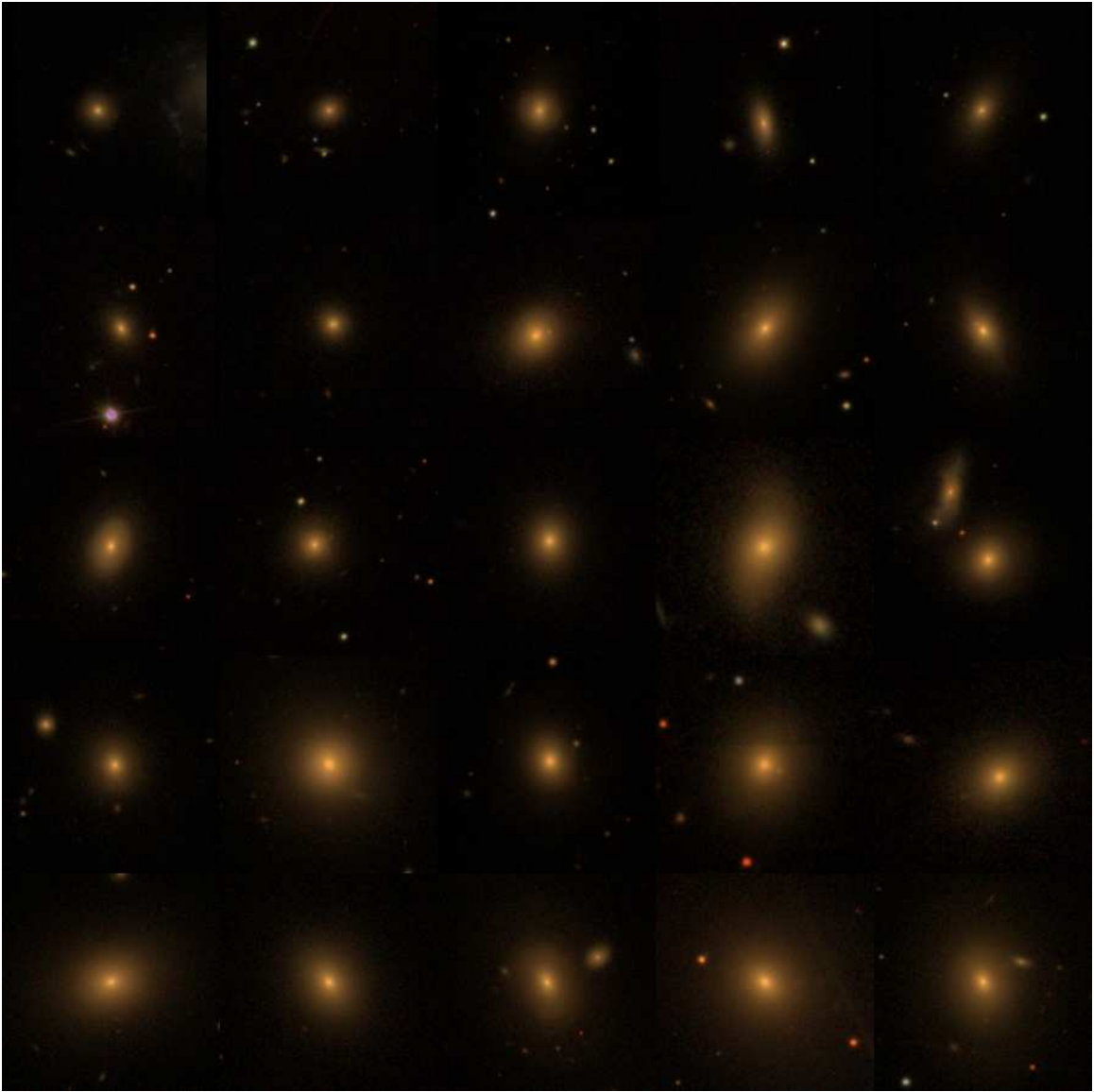}
\caption{
Randomly selected examples in the final {\tt Elliptical} sample.
We show the images, combined from images in $gri$ bands,
of 25 galaxies in ascending order of velocity dispersion ($\sigma$)
from $70~\kms$ to $325~\kms$, from top to bottom and from left to right.
}
\label{example}
\end{figure}

\begin{figure}
\epsscale{1.0}
\plotone{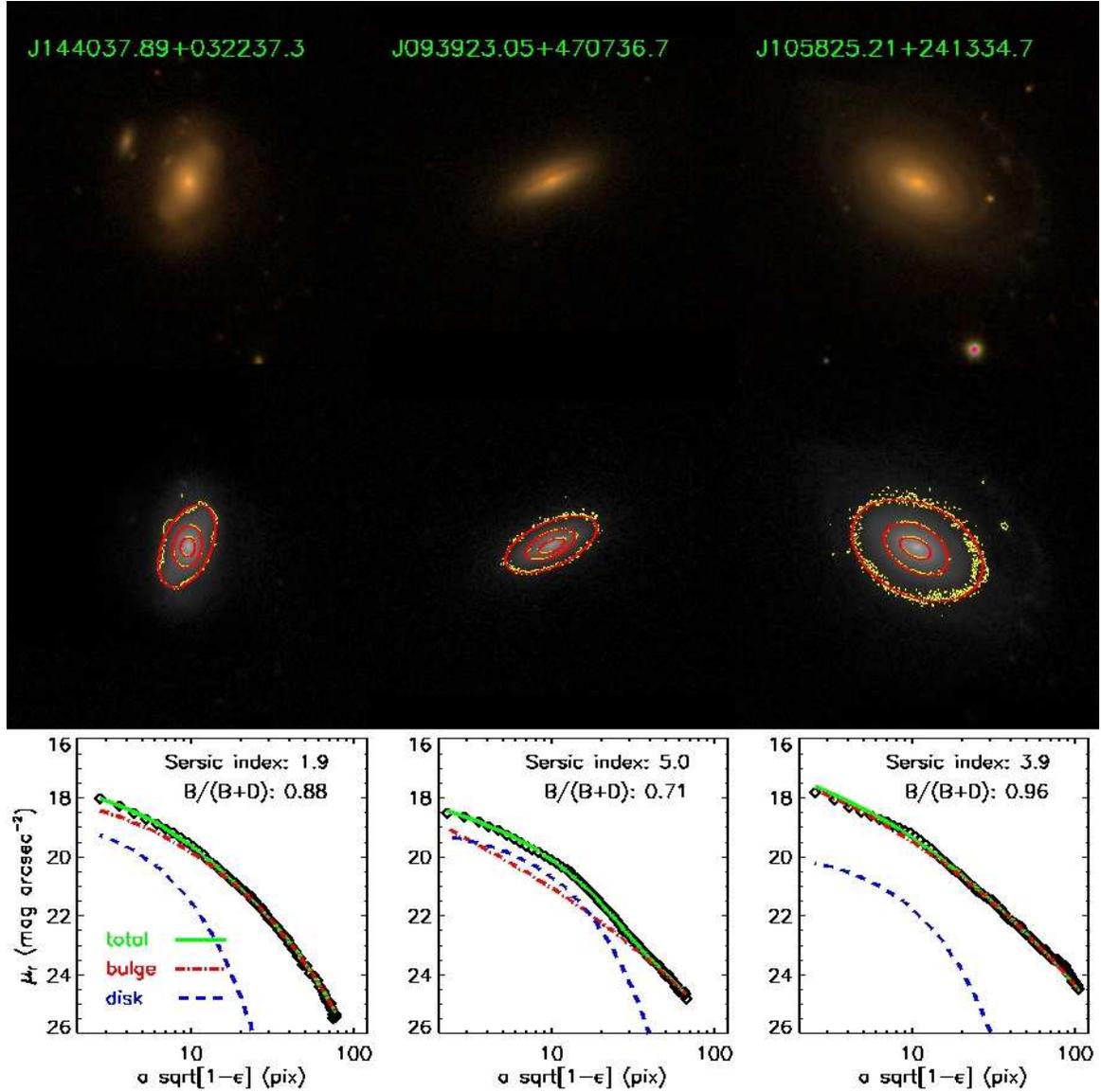}
\caption{
Examples of bulge-dominated galaxies that pass parameter cuts in preselection
but fail visual inspection.  From left to right, we show an SB0 galaxy with
a broad bar in the left panels, an S0 galaxy with a faint dust lane in the middle panels
and an S0 galaxy with faded spiral arms in the right panels.
{\it Top panels}: images, combined from images in $gri$ bands.
{\it Middle panels}: deblended images in $r$ band.  The yellow contours are the isophotes,
and the red lines are the ellipses determined by the {\tt Ellipse} algorithm.
{\it Bottom panels}: surface brightness profile in $r$ band,
decomposed to a \Sersic~component (the bulge, $B$)
and an exponential component (the disk, $D$).  $a$ stands for the major axis, $\epsilon$
represents the ellipticity, and $\mu_r$ indicates the surface brightness in $r$ band.
}
\label{rejection}
\end{figure}

Figure~\ref{example} presents color images of a randomly selected
subset of our elliptical galaxy sample, sorted by increasing velocity
dispersion.
Given our generous parameter cuts, however, our procedure does result
in some non-elliptical contaminants, three examples of which are shown
in Figure~\ref{rejection}.  Among the contaminants are bulge-dominated
SB0 galaxies with very small bars (e.g., Fig.~\ref{rejection}
\emph{left}), galaxies with faint dust lanes (e.g.,
Fig.~\ref{rejection} \emph{middle}), and S0 galaxies with faded spiral
structures (e.g., Fig.~\ref{rejection} \emph{right}).
Therefore, we visually examined the preselected elliptical galaxy sample and
excluded such galaxies.
Note that in Figure~\ref{example} we have already excluded such galaxies.

Our final sample ({\tt Elliptical}; see Table~\ref{sampletable})
contains $1,923$ elliptical galaxies with $\sigma>70~\kms$.  For those
without SDSS spectroscopy, we have performed the same analysis and
selected $430$ elliptical galaxies.  We refer to them as the bonus
elliptical sample ({\tt Bonus}).  Though we do not include them in the
analysis in the rest of this paper, we make them publicly available,
along with the {\tt Elliptical} sample with the SDSS 
spectroscopy.

Our selection certainly still suffers from some subjectivity and may
not be ``complete'' depending on one's definition of an
elliptical.  Because we aim for a clean sample, our visual inspection
is strict, and we likely exclude some elliptical galaxies that appear
ambiguous.  In addition, the bulge and disk decomposition
classification also possibly identifies real elliptical galaxies with
low \Sersic~index (low concentration) as disk-dominated galaxies.

For comparison, if we selected early-type galaxies in the same way as
in \citet{bernardi03a} or \citet{eisenstein03}, we would select $\sim
5000$ galaxies, including $\sim 1600$
of the galaxies in
our {\tt Elliptical} sample.  Therefore, we could select most of galaxies in
our sample with their selection method, but lenticulars and early-type
spirals would comprise about two thirds of the sample.

\section{The Elliptical sample}

Our final sample contains $1,923$ elliptical galaxies with
$\sigma>70~\kms$ and $z < 0.05$.  In this section, we present the
general properties of the sample, compare them with the parent and
red-sequence samples, and define the local environment of each galaxy.

\subsection{Nature of the red sequence}

Galaxies can be broadly divided into two groups, the red sequence and
the blue cloud, according to their broad-band color and luminosity
\citep[e.g.][]{blanton03d, kauffmann03, baldry04}.  In general terms,
red galaxies are early-types with a spheroidal surface brightness
profile, that are dominated by an old stellar population.
However, disk galaxies that
are dusty or have star formation quenched can also appear red in
optical broad-band colors. 
Thus, the red sequence in general consists of
both disk-dominated and bulge-dominated galaxies.

\begin{figure}
\epsscale{1.0}
\plotone{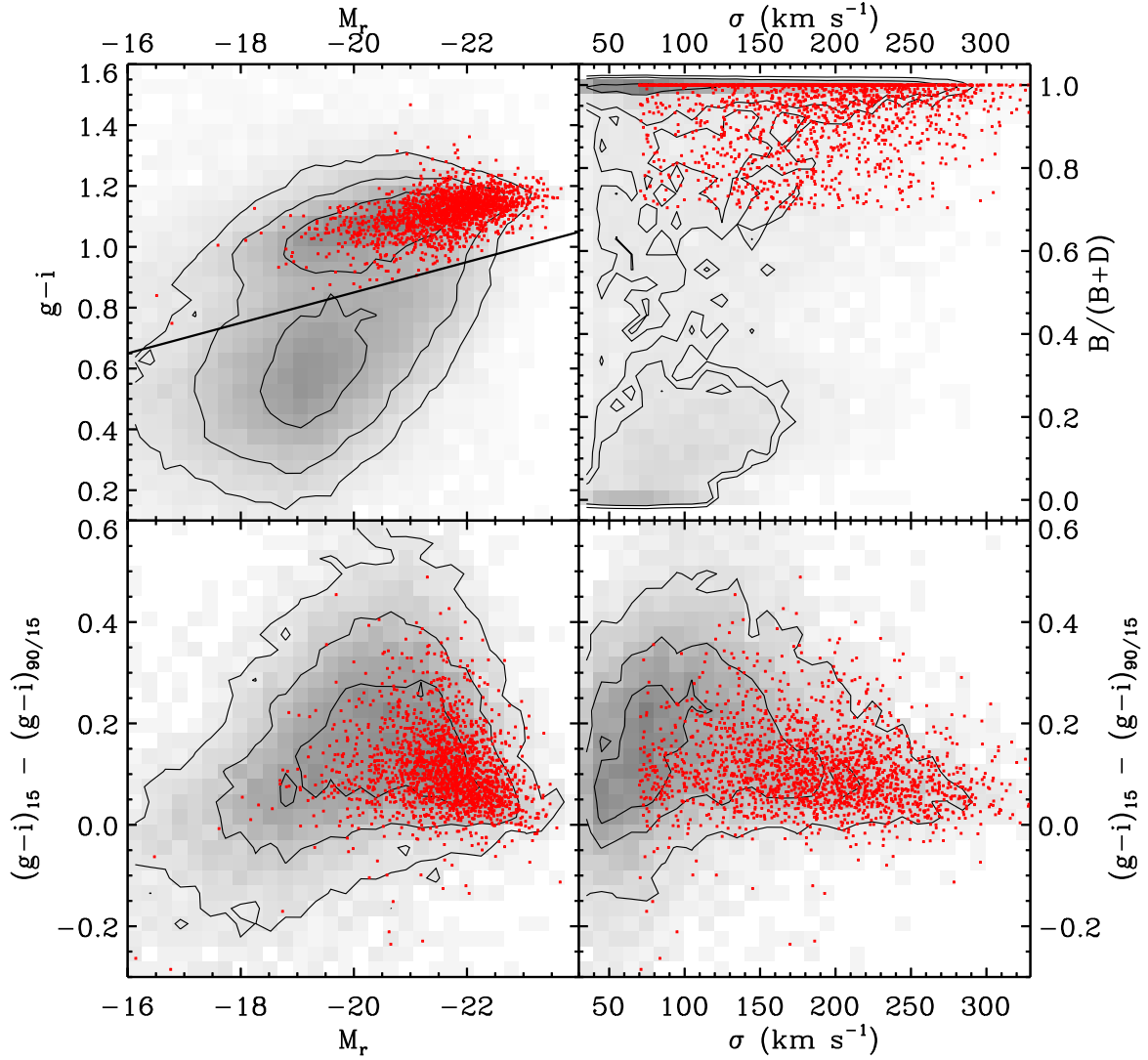}
\caption{
Broad-band properties of elliptical galaxies, compared with parent samples.
Red points indicate the galaxies in the final {\tt Elliptical} sample.
The gray scales show the density distribution of the parent samples
({\tt LowZ} in the top left panel, {\tt PhotoRS} in the bottom left panel
and {\tt SpecRS} in the right panels).
{\it Top left panel}: Color-magnitude diagram of ${g - i}$ vs. $M_r$.
The solid line shows the red sequence cut.
{\it Top right panel}: Bulge-to-total ($B/(B+D)$) ratio vs. velocity dispersion ($\sigma$).
{\it Bottom left panel}: Color gradient vs. $M_{r}$, where
the color gradient is defined as the difference between color within $15\%$
light radius ($r_{15}$) and color between the $15\%$ and $90\%$ light radii.
{\it Bottom right panel}: Color gradient vs. velocity dispersion ($\sigma$).
Elliptical galaxies dominate the bright end and form a different sequence from the disk-dominated galaxies
in the color gradient-magnitude/$\sigma$ space.
}
\label{bbp}
\end{figure}

For example, Figure \ref{bbp} shows the broad-band properties of the
{\tt Elliptical} sample (red dots) compared with the parent
samples.  The top left panel of Figure \ref{bbp} shows the full {\tt
  LowZ} sample, in which elliptical galaxies (red dots) clearly
dominate the luminous end ($\gtrsim L^\ast$) of the red sequence
\citep[see also][]{marinoni99, bundy09}.  The diagonal line corresponds
to equation (\ref{colorcut}), the color cut used to define the red
sequence in \S\ref{sec:parent}.

In Figure~\ref{bbp} ({\em top-right}) we examine the red-sequence
galaxies ({\tt SpecRS}) in more detail by plotting the bulge-to-total
ratio ($B/(B+D)$) versus $\sigma$.  Red-sequence galaxies naturally
separate into disk- and bulge-dominated groups.  Note that our cuts in
$\sigma$ and $B/(B+D)$ are visible in the distribution of the {\tt
  Elliptical} sample.  Also note that our cut on bulge-to-total ratio
tends to exclude dwarf elliptical galaxies, which often exhibit
exponential surface brightness profiles (\citealt{michard85a,
  schombert86a, prugniel97a, caon93a, graham02a, kormendy09}).

Although we do not use it for classification, we also examine the
color gradients of the galaxies in our sample.  The lower panels of Figure
\ref{bbp} show the color gradients of the red-sequence galaxies as a
function of \mr~({\tt PhotoRS}, \emph{left}) and $\sigma$ ({\tt
  SpecRS}, \emph{right}).  We define the color gradient as the color
difference between the $g-i$ color within the $15\%$ light radius and
that between the $15\%$ and $90\%$ light radii.  We find that galaxies
form two distinctly different sequences.  The central regions of most
galaxies are redder than their outer regions.  For the
disk-dominated galaxies the color gradient increases with luminosity
(\citealt{tully96, jansen00}).  Meanwhile, for the elliptical galaxies,
the color gradient decreases marginally with luminosity
(\citealt{boroson83a}; see \citealt{suh10a} for an extensive study in
a sample similar to ours).  These results suggest that the color
gradient can be used as an indicator for morphological classification
\citep[\eg,][]{park05}.
The physical cause of the color gradient is 
likely that the metallicity decreases with increasing
radius.  \citet{kuntschner10a} \citep[see also][]{spolaor09a} show that
for low-mass, fast-rotating early-type galaxies the metallicity gradient 
increases with mass, while for more massive systems the metallicity 
gradient becomes shallower, leading to the most massive systems being 
slow rotators with relatively shallow metallicity gradients.
This result is consistent with the $\sigma$/luminosity-color gradient relations
we find here.
 
\begin{figure}
\epsscale{1.0}
\plotone{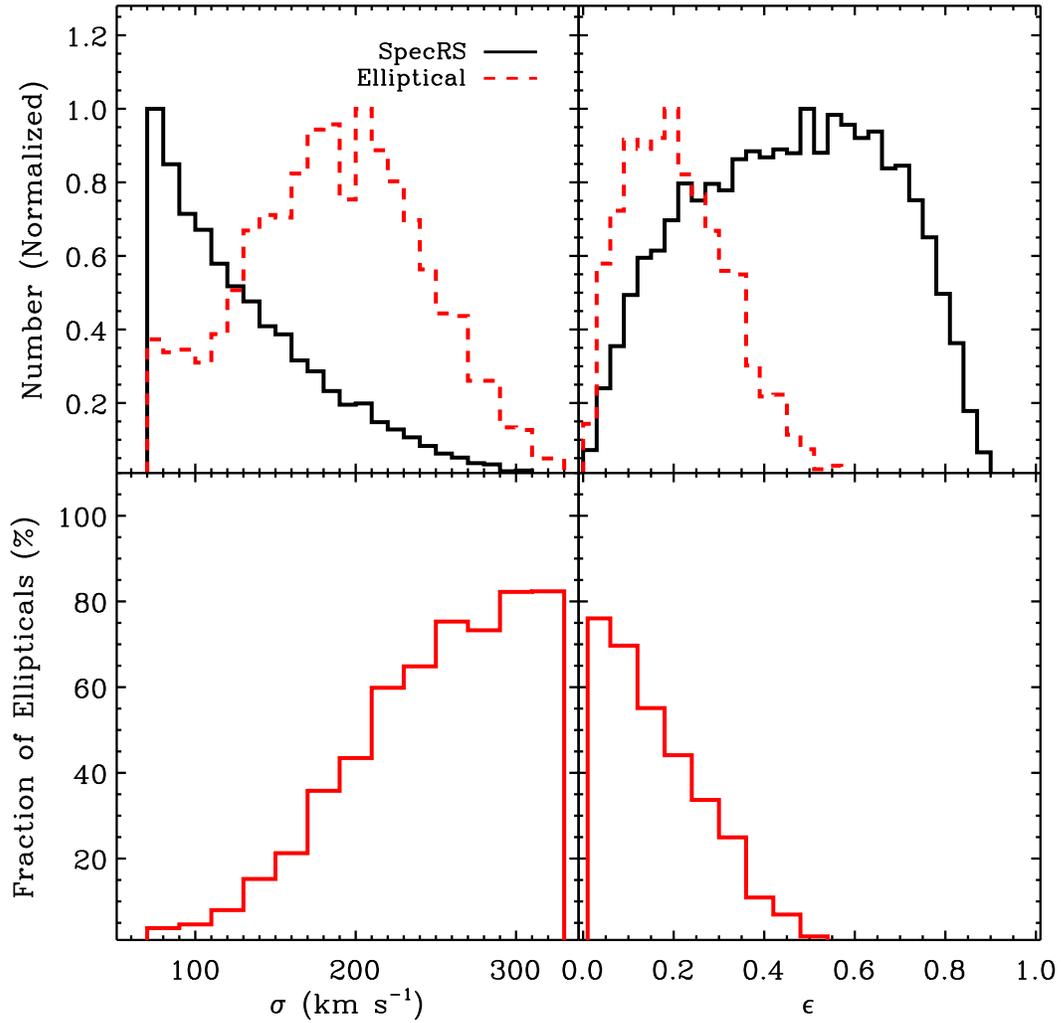}
\caption{
Distribution of velocity dispersion ($\sigma$) and ellipticity ($\epsilon$) of red-sequence galaxies and
elliptical galaxies.
{\it Top panels}: Normalized histogram of $\sigma$ and $\epsilon$.
Black solid lines indicate the distribution of the parent {\tt SpecRS} sample,
and red dashed lines represent that of the {\tt Elliptical} sample.
{\it Bottom panels}: Fraction of elliptical galaxies in the parent {\tt SpecRS} sample,
as a function of $\sigma$ and $\epsilon$.
The fraction is a strong function of $\sigma$ and $\epsilon$.
Elliptical galaxies dominate at the massive end and at the low $\epsilon$ end.
}
\label{rsmakeup}
\end{figure}

In the top panels of Figure~\ref{rsmakeup} we compare the
distributions of $\sigma$, and ellipticity ($\epsilon$) for the {\tt
  Elliptical} and {\tt SpecRS} samples.  Because $\epsilon$ in general
varies within a galaxy, we must choose a particular definition; here
we use the mean value between the $30\%$ and $60\%$ light radii.  We
find the distributions of both quantities to be very different among
the two samples.  Specifically, elliptical galaxies contain many more
massive galaxies, and have a median $\epsilon \sim 0.2$.  We also note
from Figure~\ref{rsmakeup} ({\em top-right}) that the $\epsilon<0.6$
cut we applied in \S\ref{sec:preselect} rejects very few real
elliptical galaxies.
 
The bottom panels of Figure \ref{rsmakeup} quantify the fraction of
elliptical galaxies in the {\tt SpecRS} sample as a function of
$\sigma$ and $\epsilon$.  We find that the fraction varies from
$\lesssim 10\%$ for $\sigma \lesssim 100~\kms$ and $\epsilon \gtrsim
0.5$, to $\sim 80 \%$ for $\sigma \gtrsim 300~\kms$ and $\epsilon
\lesssim 0.1$.  Note that our selection was intended to be strict;
therefore these fractions should be treated as lower limits.
Nevertheless, these results suggest that only if one restricts to the
luminous end are red-sequence galaxies typically giant elliptical
galaxies.  At fainter
luminosities ($L\lesssim L^{\ast}$), disk-dominated galaxies
constitute the bulk of the red sequence.  As we pointed out above,
part of this trend with luminosity is due to our bulge-to-total cut;
lower luminosity ellipticals in general have smaller \Sersic\ indices
\citep[\eg,][]{graham02a, kormendy09} and our procedure may start
excluding those when they become consistent with exponential profiles.
Selecting a reliable sample of fainter elliptical galaxies would
require more information than we use here.

\subsection{Local environment}

There are various ways to define the local environment of a galaxy
\citep{bernardi03a, eisenstein03, balogh04, park07, schawinski07}.
For instance, parametrizing the environment with local density 
separates the central and the outer regions of a cluster, while
using group membership can separate galaxies associated with groups
from those isolated in the field.  Here, we will compare galaxies in
groups to isolated field galaxies.

We use a Friends-of-Friends (FoF) grouping algorithm to define
groups.  We choose linking lengths to be
$0.88$~Mpc and $1.76$~Mpc ($120~\kms$) in the angular and
line-of-sight directions, respectively.  We choose the linking lengths
to be short enough to break filaments into knots, but not too short to
only select the galaxies in the center of any groups.  The
line-of-sight linking length is small compared to other
implementations in the literature \citep[\eg,][]{goto05, berlind06};
for example, \citet{berlind06} use $300~\kms$.  We explain below why
we make this choice.
 
We define the groups based on the {\tt LowZ} sample, which includes
the RC3 catalog galaxies necessary to ensure completeness at the
bright end.  However, consistently finding groups across redshift
requires a volume-limited sample.  At the faint end, the flux limit of
the SDSS spectroscopic survey is $r = 17.77$, which corresponds to
$M_r \sim -19$ at $z = 0.05$.  We therefore only include galaxies in
{\tt LowZ} brighter than $M_r = -19$ in our FoF analysis.
The resulting environmental sample ({\tt Environ}) has
$57,885$ galaxies in total.  In our {\tt Elliptical} sample, there are
only $16$ galaxies fainter than $M_r = -19$; the velocity dispersion
cut thus has provided roughly a volume-limited sample.  In the
following sections, we do not consider these $16$ faint ellipticals
when studying stellar populations.

We divide galaxies into three broad categories: rich group,
poor group, and field.  Rich-group galaxies are in groups with at
least $5$ members, poor-group galaxies are in ones with $2-4$ members,
and field galaxies are the sole members of their group.  We
note that the line-of-sight linking length ($120~\kms$) we adopt is
relatively small compared to the typical velocity dispersion of
clusters.  This choice may result in identifying a galaxy associated
with a group as an isolated galaxy.  Therefore for each rich group we
define an ellipsoid with radius $0.88$ Mpc in the angular direction
and radius $360~\kms$ in the line-of-sight direction. For each
poor group, we also define such an ellipsoid but adopting a slightly
smaller radius $240~\kms$ in the line-of-sight direction. If an
isolated galaxy lies within the ellipsoid of any group, we consider
its group classification to be ambiguous.  There are $129$ ambiguous
galaxies in total, $46$ of them with possible association with rich
groups and $83$ with poor groups. Including them either in the
sample of field galaxies or group galaxies does not affect any of
our results; therefore, we exclude them from further analysis.

In \S\ref{sec:analysis} we study the environmental dependence of the
stellar populations of our sample.  However,
it is important to exclude low signal-to-noise ratio ($S/N$) 
spectra and strong AGN.  We thus
will not consider galaxies with spectra of median $S/N$ (per pixel) less than
$10$.  To exclude strong AGN, we use a standard emission
line diagnostic diagram (\citealt{baldwin81}, BPT).  For galaxies with
emission lines \Hb, \oiiilam, \Ha~and \niilam~of $S/N>5$, we exclude
them if they appear to the right of the demarcation line defined by
\citet{kauffmann03} in the BPT diagram.  We exclude $39$ strong AGN in
this process, which we have verified does not affect the
analysis presented below.

To summarize, after excluding $59$ galaxies with $M_r > -19$
(16), spectra of low $S/N$ (4), strong AGN (39), and $129$ galaxies
with ambiguous group association, out final {\tt Elliptical} sample
consists of $347$ field elliptical galaxies, $682$ poor-group
galaxies, and $706$ rich-group galaxies.
 
\begin{figure}
\epsscale{1.0}
\plotone{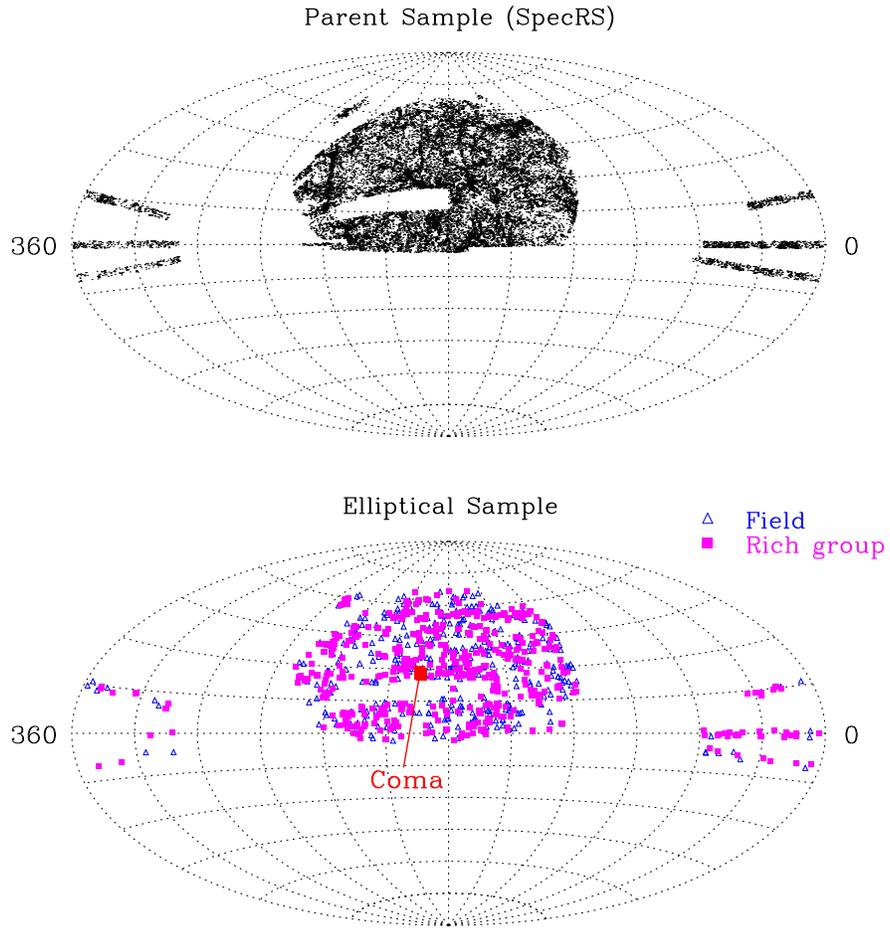}
\caption{
Angular distribution of the parent {\tt SpecRS} sample and
part of the {\tt Elliptical} sample in equatorial coordinates.
In the upper panel, we show the galaxies in {\tt SpecRS} as black points.
We show the elliptical galaxies in the bottom panel.
Blue open triangles represent field galaxies,
magenta filled squares represent rich-group galaxies.
We also show the position of the Coma cluster in the bottom panel.
}
\label{environment}
\end{figure}

Figure \ref{environment} shows the angular distribution of the {\tt
  Elliptical} sample.  In the top panel, we show the distribution of
the red-sequence galaxies in the parent {\tt SpecRS} sample.  In the
bottom panel, we show the distribution of part of the {\tt Elliptical}
sample, where we represent field galaxies as blue open triangles, and
rich-group galaxies as magenta small solid squares.  We also show the
position of the largest group, the Coma cluster as a large red solid
square.  
 
\section{Stellar populations of elliptical galaxies}\label{sec:analysis}

In this section, we study the stellar population of our {\tt Elliptical} 
sample as a function of $\sigma$ and environment with average optical spectra
and Lick indices.

\subsection{Average spectra}

\subsubsection{Stacking method}

We stack the spectra in various samples of velocity
dispersion and group classification, as described throughout.  There
are three versions of stacks we make here: unsmoothed, smoothed, and
uniform stacks.

For each sample, we create at least 
the unsmoothed and smoothed average spectra.  In the
unsmoothed version, we do not smooth individual spectra before
averaging them.  In the smoothed version, we first smooth each
spectrum to the largest $\sigma$ in that sample.  The unsmoothed
version allows us to see more details that are washed out in the
smoothed version, while the smoothed version allows us to
compare samples across different environments consistently.

For samples in different $\sigma$ bins, we also calculate a uniform
version by smoothing each individual spectrum to $325~\kms$, to match
the largest $\sigma$ in the whole sample.  These stacked spectra
allows us to compare different samples as a function of
velocity dispersion consistently.  We use this uniform version to
measure the Lick indices for the average spectra in the following
sections ($\S 4.2.3$ and $\S 4.2.4$).

\begin{figure}
\epsscale{1.0}
\plotone{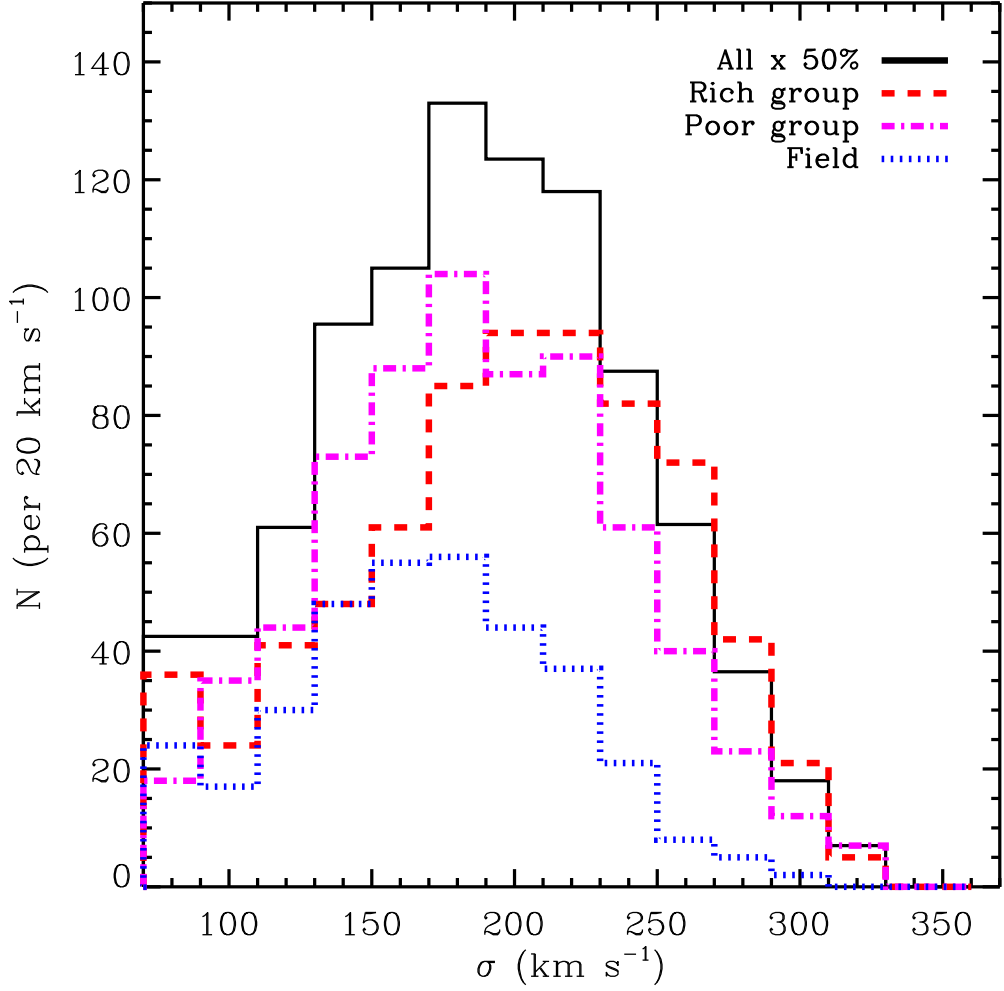}
\caption{
Distribution of velocity dispersion ($\sigma$) of elliptical galaxies as a function
of environment.
We show the distribution of number of galaxies per $20~\kms$.
For clarity, we have multiplied the number of the whole sample by $50\%$.
The field galaxies tend to have lower $\sigma$.
The median $\sigma$ for field, poor-group, and rich-group galaxies are
$170$, $186$, and $203~\kms$, respectively.
}
\label{vdispdist}
\end{figure}

Figure \ref{vdispdist} shows the velocity dispersion distribution as a
function of group classification.  The galaxies in rich groups tend to
have higher $\sigma$ than those in low-density environments.  As we
show explicitly below, and is already well known, the properties of
elliptical galaxies depend strongly on $\sigma$ and luminosity.  For
instance, the color of elliptical galaxies is redder at higher
luminosity and larger $\sigma$ (Figure \ref{bbp}; \eg,
\citealt{baum59}).  

Thus, to make a direct comparison between samples in different
environments, we need to ensure that we compare samples with
the same $\sigma$ distribution --- otherwise we will simply be
measuring the dependence on $\sigma$ itself, which dwarfs all other
effects.  To do so, we weight the galaxies as a function of $\sigma$
in order to achieve the same effective $\sigma$ distribution for each
subsample.  Within each environmental subsample, for a galaxy with a
certain $\sigma$, we determine the number of galaxies within $\Delta
\sigma = \pm 10~\kms$ in both the full sample and in the subsample.
We then weight each spectrum in the stack by the number in the full
sample divided by the number in the subsample.  In this way, the
different environmental samples have the same effective $\sigma$
distribution.  This procedure allows us to separate the environmental
dependence from the $\sigma$ dependence.

The SDSS spectra are observed between wavelengths of $\sim 3800$
\AA~and $\sim 9200$ \AA.  Because our redshift limit is $z=0.05$, for
consistency we calculate the average spectra in the rest-frame
wavelength range available for all galaxies, from $3800$ \AA~to $8800$
\AA.  Before stacking the spectra, we also normalize each spectrum to
its mean flux between $5200$ \AA~and $5800$ \AA, where the spectra are
relatively flat.

The spectrophotometry\footnote{\tt
http://www.sdss.org/dr6/algorithms/spectrophotometry.html} in the
SDSS is calibrated with F8 subdwarfs and is accurate at the few
percent level.  In any case, our differential results between
different environments and velocity dispersion samples are largely
unaffected by systematic errors, unless their strength correlates with
those parameters, which is unlikely.

\subsubsection{Velocity dispersion dependence}

\begin{figure}
\epsscale{0.85}
\plotone{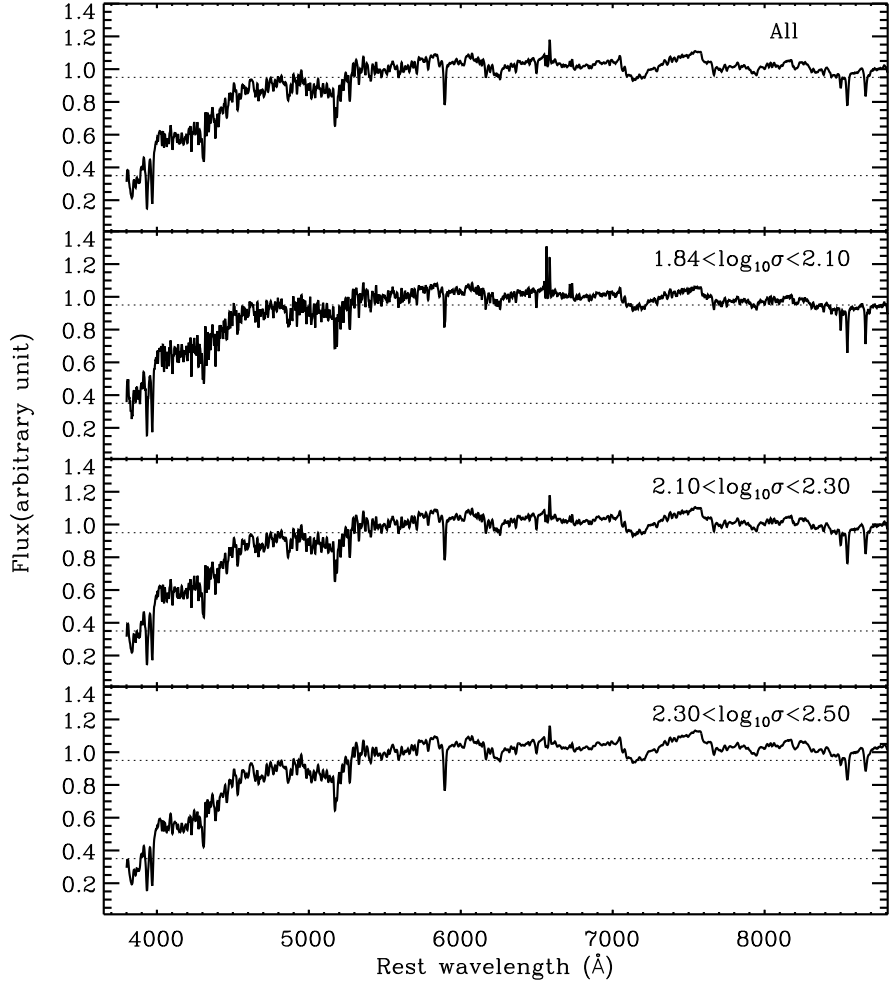}
\caption{
Average spectra (unsmoothed version) of elliptical galaxies as a function of velocity dispersion ($\sigma$).
Before stacking each spectrum, we normalize it to the mean flux between $5200$ \AA~and $5800$ \AA~where
the spectrum is relatively flat.
When stacking the spectra, we mulptiply each spectrum in each subsample
by a weight factor that is the ratio of the number of the galaxies
in the whole sample to that in the subsample at the same velocity dispersion ($\sigma$, Figure \ref{vdispdist}).
This ensures that we are comparing spectra in all the subsamples, e.g., in different environments,
with the same effective $\sigma$ distribution.
We also calculate the jackknife errors by dividing each subsample into 10 sub-subsamples
with equal number of galaxies.
The typical errors are smaller than $1\%$ and we do not show them in this plot.
We define the $\sigma$ bins to be of roughly equal size in log space.
The two dotted lines are the same in each panel ($0.35$, $0.95$) to guide the eye.
The average spectra at lower $\sigma$ are apparently bluer.
}
\label{vdispavgspec}
\end{figure}

Figure \ref{vdispavgspec} shows the average unsmoothed spectra
as a function of $\sigma$.  We overplot two horizontal lines at $0.95$
and $0.35$ in each panel to guide the eye.  We define the $\sigma$
bins to be roughly equal size in log space.  The three bins (in
$\log_{10}\sigma$) are [$1.84$, $2.10$], [$2.10$, $2.30$] and [$2.30$,
  $2.50$].  The median $\sigma$ in each bin are $100$, $167$ and
$232~\kms$, respectively.  The average spectra at all $\sigma$ are
typical of an old stellar population, with strong absorption features
such as the $4000$~\AA~break (mainly caused by the CaII H and K
lines), the G-band at $4300$~\AA, the $5180$~\AA~MgI and H feature,
the $5890$~\AA~Na I and the $7200$~\AA~TiO lines.  Though we have
excluded strong AGN when calculating the average spectra, weak
emission features such as \Ha~are still visible, though at a fairly
weak level.  They can either be caused by low-level star formation or
weak AGN activity.  Line ratio diagnostics show that some of the
elliptical galaxies with emission are LINER-like \citep[see also,
\eg,][]{yan06, graves07} while others show signs of
ongoing star formation.

\begin{figure}
\epsscale{1.0}
\plotone{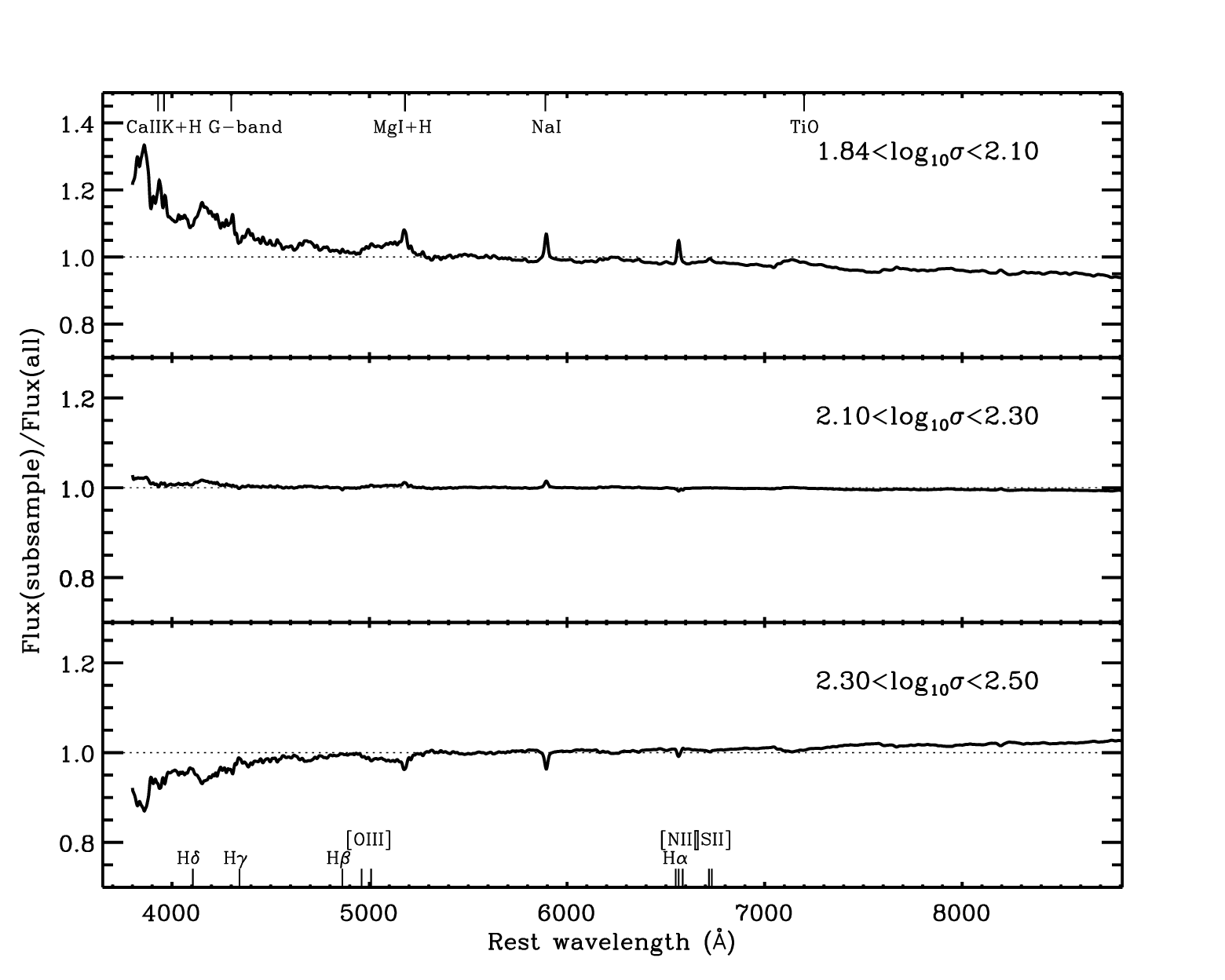}
\caption{
Average spectra (uniform version) as a function of velocity dispersion ($\sigma$).
We show the ratio of the average spectra of all elliptical galaxies in each $\sigma$ (in $\kms$) bin
to that of all elliptical galaxies.
}
\label{vdispavgspecratio}
\end{figure}

Not surprisingly, we see a clear dependence on velocity dispersion of
the average spectra (which is why correcting for the $\sigma$
distribution is so important when studying environmental effects).
The average spectra in lower $\sigma$ bins are relatively bluer than
those in higher $\sigma$ bins, which is consistent with the
color-magnitude relation \citep[\eg,][ among many others]
{visvanathan77, strateva01, hogg04, baldry04, balogh04}.  
We examine this result in more detail in Figure 
\ref{vdispavgspecratio}, where we
show the ratio of the average spectra (uniform version) of all
galaxies within different $\sigma$ bins to that of the whole sample.
We find that lower-$\sigma$ galaxies exhibit a bluer continuum,
weaker metal absorption features and slightly stronger emission
lines (\eg, \Ha).

However, the cause of the bluer continuum for elliptical galaxies at
lower $\sigma$ is less clear.  The primary difficulty in
interpretation is the well-known age-metallicity degeneracy in stellar
population analysis
\citep[\eg,][]{faber72, faber73, oconnell80, rose85, renzini86,
worthey94, bruzual03}.  Both increasing age and increasing metallicity
have extremely similar effects on the spectral energy distributions
(SEDs).  This similarity means that either a younger population or a
lower metallicity population (or some combination) can be responsible
for the bluer continua of elliptical galaxies at lower $\sigma$.
Furthermore, the presence of metal-rich, old blue horizontal branch
stars \citep[\eg,][]{sweigart87, lee90, lee94} can mimic a younger
stellar population and introduce more flux at the blue end as well
\citep[\eg,][]{yi97, maraston00, lee00, conroy10a}.  Blue stragglers
\citep[\eg,][]{bailyn95, brown05} that extend blueward and 
brightward of the main-sequence turnoff point may also have a non-negligible
effect on integrated spectra \citep[\eg,][]{xin05, conroy10a}.
Stronger AGN activity or less dust extinction can also cause the same
effect.

Because of these effects, we cannot be certain what is responsible for
the velocity dispersion dependence in the average spectra.  However,
the stronger emission lines, part of which certainly is due to star
formation activity, and the weaker metal absorption features suggest
that both younger ages and lower metallicities are responsible for
the bluer continuum at lower $\sigma$.

\subsubsection{Environmental dependence}

\begin{figure}
\epsscale{1.0}
\plotone{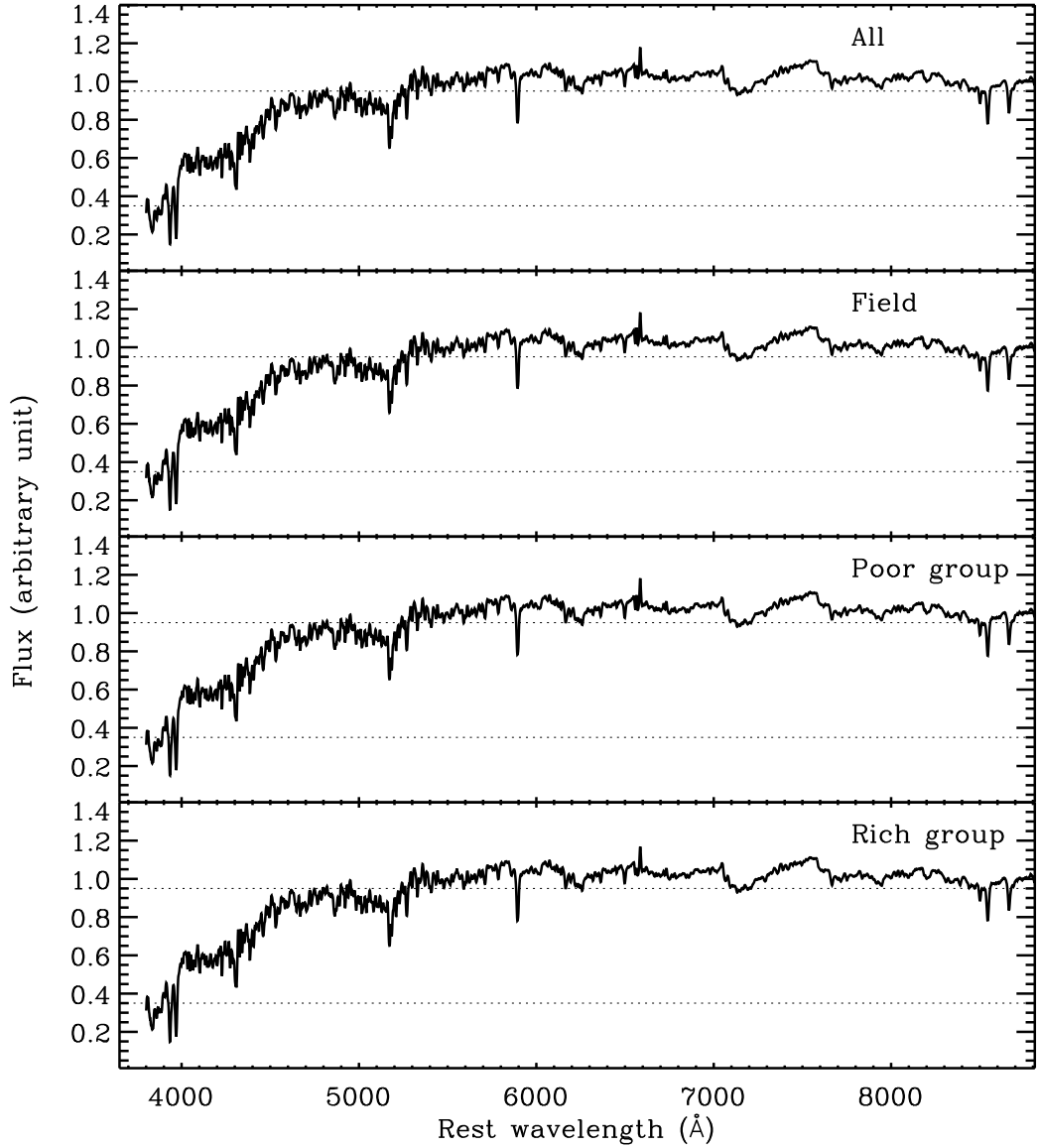}
\caption{
Average spectra (unsmoothed version) of elliptical galaxies as a function of environment.
They look strikingly similar to each other.
However, small differences do exist when we take a closer look in Figure \ref{avgspecratio}.
Note it is important to control for $\sigma$ when comparing samples in different
environments (See Figure \ref{vdispavgspec} and \ref{vdispavgspecratio}).
}
\label{avgspec}
\end{figure}

Figure \ref{avgspec} shows the average unsmoothed spectra as
a function of environment.  Recall, in these stacks we have weighted
the spectra so that the effective $\sigma$ distributions are the same;
otherwise, stronger trends would be visible due to the correlation of
$\sigma$ with environment.  The average spectra for all the
environmental samples look strikingly similar to each other.  These
spectra reveal that old stellar populations dominate the stellar
components in elliptical galaxies in all environments.  

\begin{figure}
\epsscale{1.0}
\plotone{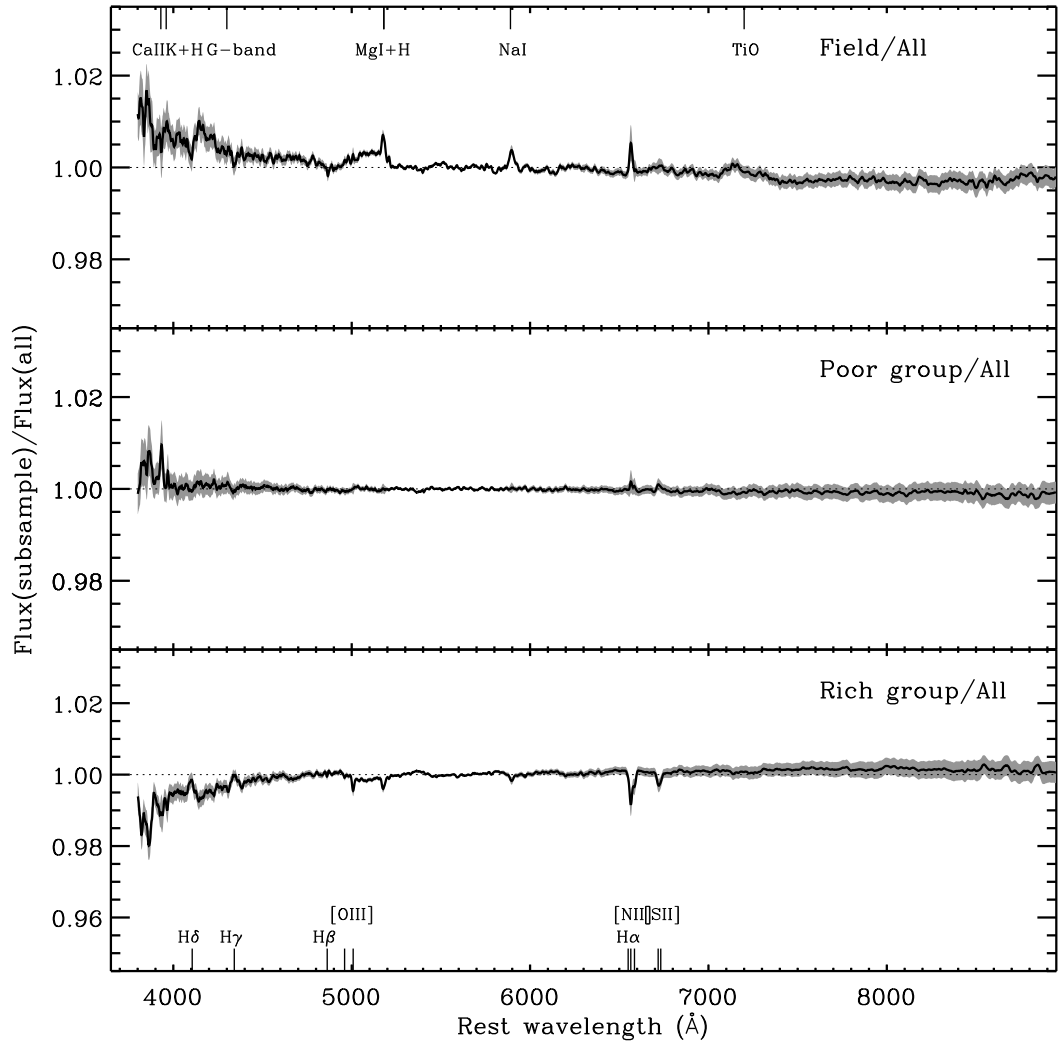}
\caption{
Comparison of average spectra (smoothed version) of elliptical galaxies in different environments.
We show the ratio of the average spectra of the subsamples to that of
the whole sample.  We calculate the jackknife errors by dividing
each subsample into 10 sub-subsamples with equal number of elliptical galaxies
and show them separately with gray scales.
The field galaxies have a bluer continuum and stronger Balmer and nebular emission lines,
but by only $\sim1$ percent compared to the whole sample.
The rich-group galaxies, on the contrary, have a redder continuum and weaker Balmer
and nebular emission lines, but also by only $\sim 1$ percent.
}
\label{avgspecratio}
\end{figure}

Although by eye the average spectra for all the samples look exactly
the same, careful inspection shows subtle variations.  In Figure
\ref{avgspecratio}, we present the ratio of 
the average spectra (smoothed version) of
different samples in different environments to that of the whole
sample.  We find weak but significant differences between the average
spectra in different environments.  Compared to the average spectra of
the whole sample, the field galaxies have a bluer continuum,
especially at wavelength $\lesssim 4000$~\AA, by $\sim 1$ percent, and
have stronger (but still weak) emission lines of \Hb, \oiiidoublet,
\Ha, \niidoublet~and \siidoublet~etc.  The rich-group galaxies, on the
other hand, have a redder continuum and have less flux at the blue end
by $\sim 1$ percent, and have weaker emission lines.

\begin{figure}
\epsscale{1.0}
\plotone{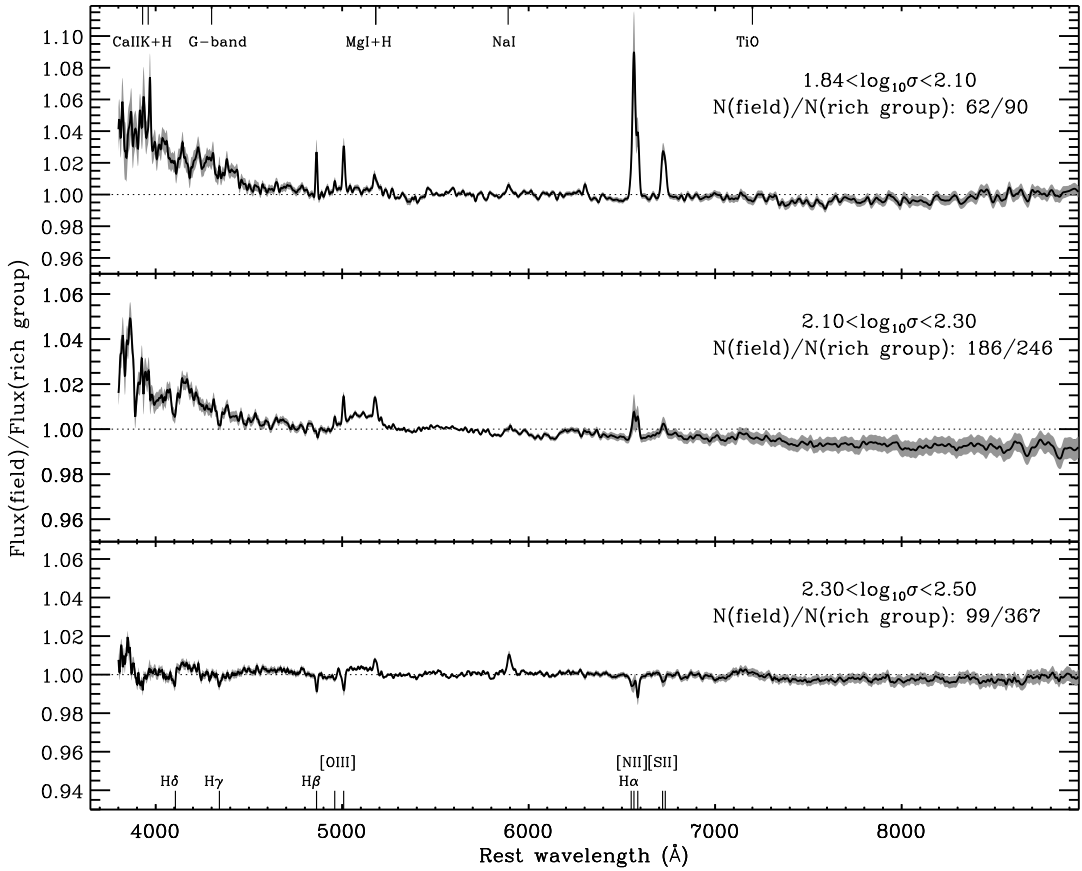}
\caption{ Comparison of average spectra (uniform version) of field
elliptical galaxies with that of rich-group elliptical galaxies as a
function of velocity dispersion ($\sigma$).  N(field) indicates the
number of field elliptical galaxies, and N(rich group) indicates
that of rich-group elliptical galaxies in each $\sigma$ bin.  The
gray scales shown are jackknife errors.  Compared to rich-group
galaxies, field galaxies have a stronger continuum and stronger
Balmer and nebular emission lines.  The environmental dependence
appears to be strong at low $\sigma$ and vanishes in the highest
$\sigma$ bin.}
\label{fieldtogroup}
\end{figure}

To further inspect the environmental dependence, we compare galaxies
in different environments in each bin of velocity dispersion.  In
Figure \ref{fieldtogroup}, we show for each $\sigma$ bin the ratio of
the average spectrum (uniform version) of the field galaxies to that
of the rich-group galaxies.  Interestingly, we see that in all three
bins, the average spectra of the field galaxies are bluer than their
rich-group counterparts.  However, the weak environmental dependence
appears to be a strong function of $\sigma$.  In the lowest $\sigma$
bin, the field galaxies have about 5\% more flux at the blue end than
the rich-group galaxies.  In the highest $\sigma$ bin, the field
galaxies have only about $\sim 1$ percent more flux at the blue end
than the rich-group galaxies.  The stronger emission lines in the low-$\sigma$
field galaxies also appear to vanish for the high-$\sigma$ field galaxies.

As was the case for the dependence on $\sigma$, the cause of the weak
environmental dependence of the average spectra is not absolutely
clear.  However, the difference in emission lines suggests that
stronger (but still low-level) recent star formation could be the
cause of the bluer spectra in the field galaxies.  In the next subsection,
we examine this hypothesis more carefully.

\subsubsection{Discussion: stronger low-level recent star formation
  activity at lower velocity dispersion and in the field?}\label{sec:losfr} 

Physical degeneracies make it difficult to derive accurate
stellar population properties from the spectra of galaxies with
certainty --- such attempts suffer from the age-metallicity
degeneracy, the possibility of poorly understood blue
horizontal branch and/or blue straggler populations, and other
effects.  In this section we take a simplified view and evaluate what
kind of young stellar population would be necessary to explain the
observed trend of SED with environment.  

First, we assume that elliptical galaxies consist of an old
base stellar population and a small frosting of young stars from
low-level recent star formation (\eg, \citealt{trager00b, gebhardt03};
S07).  Furthermore, we assume that the strength of the recent star
formation is the main factor that causes the $\sigma$ dependence and
the environmental dependence.

Under these assumptions, we fit each individual spectrum with a model
consisting of a simple combination of old and young stellar
populations.  The old components consist of $3$ old stellar
populations with ages and metallicities as follows: $13.5$ Gyr and
solar abundance ($\mathrm{Z_\odot}$); $15$ Gyr and $0.4$
$\mathrm{Z_\odot}$; and $12$ Gyr and $2.5$ $\mathrm{Z_\odot}$.  The
young components consist of $3$ young templates with ages and
metallicities as follows: $0.9$ Gyr and $\mathrm{Z_\odot}$, $1$ Gyr
and $0.4$ $\mathrm{Z_\odot}$, and $0.8$ Gyr and $2.5$
$\mathrm{Z_\odot}$ respectively.  All these templates are from
\citet{bruzual03} SSP models with the Padova (1994) library of stellar
evolution tracks \citep[e.g.,][]{alongi93, bressan93, fagotto94a,
  fagotto94b, girardi96} 
and the \citet{chabrier03} initial mass function from $0.1$ to
$100~M_\odot$.  For each individual galaxy spectrum, we mask out the
emission line regions and fit the continuum to a nonnegative linear
combination of these six templates.

We evaluate how much recent star formation there is by comparing the
resulting mass fraction $f_y$ contained in the young templates in the
fits.  We choose $f_y=2\%$ to be the minimum fraction of the young
component at which we consider the recent star formation to be
significant.  This choice allows a convenient comparison with previous
work using GALEX NUV photometry.  For reference, $f_y = 2\%$ gives the
color NUV$-r \sim 5.4$, which is the color cut adopted by
\citet{schawinski07} to indicate recent star formation. 

\begin{figure}
\epsscale{1.0}
\plotone{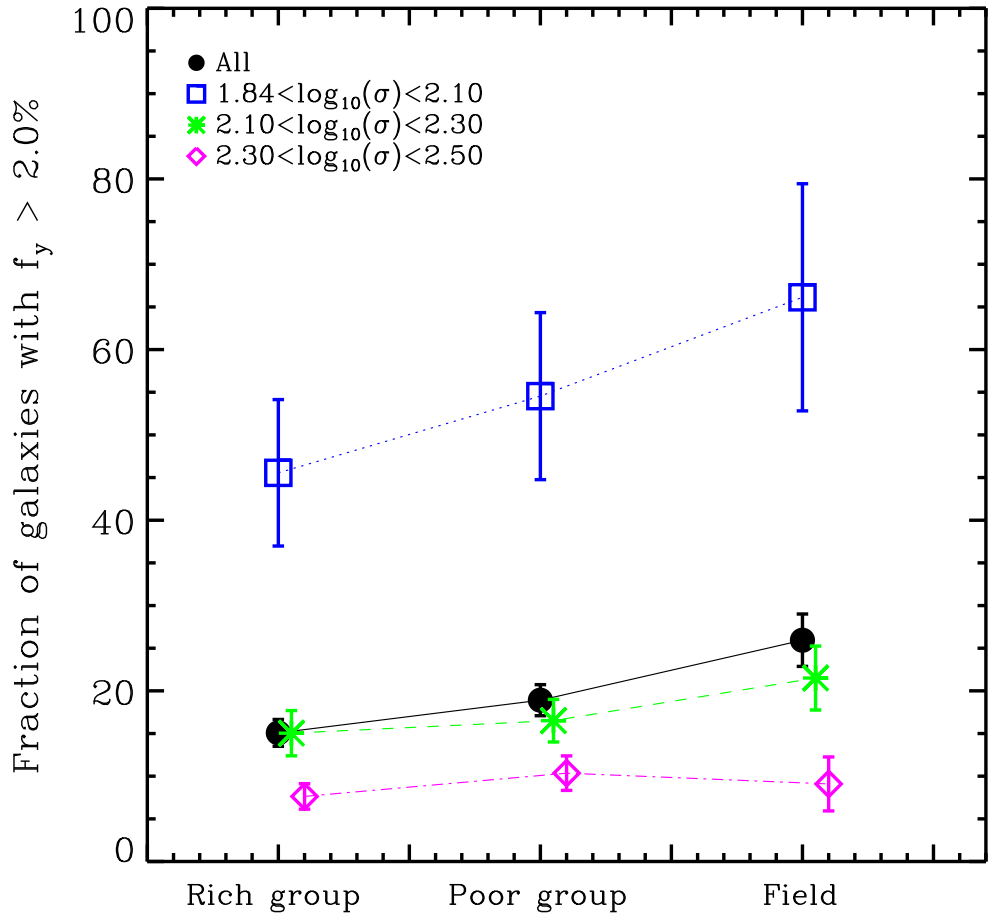}
\caption{
Fraction of galaxies with the mass fraction of the young component $f_y > 2.0\%$ as a function of environment
and velocity dispersion ($\sigma$).
We fit each spectrum with a two-component model (young+old, $\S 4.1.4$).
$f_y$ is the mass fraction of the young component.  $f_y=2.0\%$
gives NUV$-r \sim 5.4$, which is the cut \citet{schawinski07} adopted
to indicate recent star formation.
This is based on the assumptions that age is the dominant factor that shapes SEDs, and
elliptical galaxies consist of an old base stellar population and a small fraction of
young stars (the frosting model).
The fraction of galaxies with significant young stellar populations is a strong
function of $\sigma$ and environment, consistent with the average spectra.
The fraction is higher at low $\sigma$ and in the field.
The environmental dependence is strong at low $\sigma$ and vanishes at high $\sigma$.
}
\label{rsffit}
\end{figure}

In Figure \ref{rsffit}, we show that the fraction of galaxies
with significant recent star formation ($f_y > 2\%$) is a
strong function of both environment and velocity dispersion.
Overall, about $20\%$ of the elliptical galaxies have more than $2\%$
of the mass contributed from the young components.  In the lowest
$\sigma$ bin, about half of the elliptical galaxies have more than
$2\%$ young components.  Meanwhile, in the highest $\sigma$ bin, only
$\lesssim 10\%$ have $f_y>2\%$.  The fraction of galaxies with recent
star formation also appears to be larger for the field galaxies than
the rich-group galaxies, by $\sim 30\%$ in the lowest $\sigma$ bin.
As suggested by the average spectra, the environmental difference
vanishes in the highest $\sigma$ bin (Figure \ref{fieldtogroup}).

Interestingly, our results agree reasonably well with previous work
\citep[\eg,][]{schawinski07}.  Using GALEX NUV photometry of a sample
of $839$ bright early-type galaxies with $M_r < -21.5$, most of which
are elliptical galaxies, selected in the SDSS in the redshift range
$0.05 < z < 0.10$, \citet{schawinski07} find that $30\% \pm 3\%$ of
massive early-type galaxies have NUV$-r$ color bluer than $5.4$.  They
also show that the fraction is $25\%$ higher in the lowest-density
environment, which is defined based on the number and distance of
galaxies around each object (the density).  Despite the differences in
the definition of environment, as well as the different samples and
data used, our results are in fair agreement with each other.
However, they did not see the luminosity/mass dependence that we see
here, probably because of their smaller sample size and luminosity
range.

Using a sample of early-type galaxies at $0.05<z<0.06$
from MOSES, \citet{thomas10} find that the distribution 
of ages is bimodal with a strong peak at old ages and a secondary peak
$\sim 2.5$~Gyr younger containing $\sim10$ percent of 
the objects.  Interestingly, they find that the fraction of 
young galaxies increases with decreasing galaxy mass and 
decreasing environmental density. They also find that the environmental 
dependence is most pronounced at low $\sigma$ and disappears at high 
$\sigma$ (see Fig.~$8$ in their paper). 
Despite the considerable differences in sample
selection and analysis,
our results are remarkably consistent with one another.  
\citet{thomas10} further show that the young galaxies
have lower \aFe~ratios than average and most of them show signs of
ongoing star formation through their emission line spectra, in line
with our conclusion that recent star formation is likely the cause of
the environmental dependence at low $\sigma$.

We emphasize that we have assumed that age is the dominant factor
that determines the shape of the optical SED.
The dependence on velocity dispersion and
environment could also be caused by differences in metallicity,
differences in the contribution from blue horizontal branch stars or
blue stragglers, or perhaps something else.  To accurately interpret
the spectra is an extremely difficult task due to the lack of
calibration data (see the recent reviews by \citealt{conroy09,
conroy10a, conroy10b} and references therein).  

Nevertheless, recent detailed studies show that star formation
indeed occurs in many early-type galaxies \citep{yi05, rich05,
  kaviraj07b, schawinski07, temi09, young09, salim10}.
\citet{salim10}, using high resolution far-UV (FUV) imaging with the
Solar Blind Channel of the Advanced Camera for Surveys onboard the
HST, show that for most early-type galaxies with recent star
formation, it takes the form of wide or concentric UV rings.  The
SAURON team (\citealt{shapiro10}, see also \citealt{kuntschner10a}),
find that star formation in early-type galaxies happens exclusively in
fast-rotating systems and occurs in two distinct modes.  In one mode,
star formation is a diffuse process, corresponding to widespread young
stellar populations and high molecular gas content, possibly due to
(mostly minor) mergers \citep[see also][]{kaviraj09}.  In the other
mode, star formation is concentrated into well-defined disc or ring
morphologies, which may have been caused by rejuvenations within
previously quiescent stellar systems.  Indeed,
\citet{kuntschner10a} find that the most extreme cases of
post-starburst early-type galaxies, with SSP-equivalent ages of
$\lesssim3$ Gyr, frequently show signs of residual
star formation and are generally low-mass systems.
Therefore, variations in recent star formation is a plausible cause
for both strong velocity dispersion dependence and weak but
significant environmental dependence we see here.

\subsection{Lick indices}
 
The age-metallicity degeneracy has haunted stellar population analysis
for decades.  Nevertheless, a promising approach to breaking it
remains the combined use of multiple absorption-line indices
\citep[e.g.,][]{faber85, gonzalez93, worthey94, worthey97, trager98}.
In this and the next section
we measure the absorption-line indices of our
spectra and compare them with state-of-the-art SSP models in
order to infer their ages, metallicities, and $\alpha-$enhancements.

\subsubsection{Lick indices of smoothed flux-calibrated
  spectra}\label{sec:licksmoothed} 

Lick indices measure the absorption-line strength of features in
the SEDs and are widely used in studying stellar populations.  The
standard Lick indices are measured on low-resolution spectra that are
observed with the Lick IDS instrument and not flux-calibrated.  The
SDSS did not observe the bright stars in the Lick library; therefore we
cannot match the fluxing and calibration of the Lick system.
Therefore we attempt an intermediate step by measuring the Lick
indices of smoothed flux-calibrated SDSS spectra.

The bandpasses of the Lick indices studied here are defined in Table 1
of \citet{worthey94} and Table 1 of \citet{worthey97}.  To calculate
the Lick indices, we first smooth each spectrum to a resolution that
is equivalent to a velocity dispersion of $\sigma = 325~\kms$, as
observed under the SDSS resolution ($69~\kms$), to match the largest
$\sigma$ of our sample.  We then calculate the Lick indices with a
modified version of the {\tt Lick\_EW} routine in {\tt
Ez\_Ages}\footnote{\tt
http://www.ucolick.org/$\sim$graves/EZ\_Ages.html} package developed
by \cite{graves08}.  The {\tt Lick\_EW} routine reports the errors of
each Lick index calculated in the way suggested by
\citet{cardiel98}.  

The combined resolution of $\sigma = 325~\kms$ and the instrumental
resolution of the SDSS ($69~\kms$) is at lower resolution than 
the Lick IDS instrument resolution at which
the Lick indices are defined.  We therefore need to correct the
measurements back to the effective Lick IDS resolution.  To do so, we
use the correction factors given by S07 in their Table 46, which are
for a $14.1$ Gyr old stellar population with solar metallicity.  The
factors vary at the few percent level for stellar populations with
different ages and metallicities; we ignore those tiny differences
here.  We also note that these empirical corrections for $\sigma$ may
have large uncertainties \citep[e.g.,][]{trager98}, which
unfortunately are hard to quantify and we also ignore here.

In the end, the Lick indices measured in this way are matched in
resolution, but still not with the fluxing of the Lick IDS system.
S07 gives empirical zero-point corrections (Table $1$ in their paper).
When necessary, we adopt these corrections to compare flux-calibrated
Lick indices with standard ones.  These corrections have very large
uncertainties; therefore we remind our readers that any comparison of
our measurements with the standard Lick indices should be taken with
some caution.

\subsubsection{Emission line infill correction}

One of the most challenging issues in Lick index measurements is
emission line contamination.  The Balmer absorption features are
contaminated by emission from ionized gas, either from star formation,
AGN activity, or interstellar shocks. This emission line
infill severely affects the measurement of the Balmer indices, which
are thought to be the best age indicators.  For example, a
$0.1$~\AA~contamination to the absorption-line equivalent
width (EW) of \Hb{} from emission translates to an SSP age $\sim1$ Gyr
older (\eg, S07).

We correct for emission-line contamination using \oiiilam,
which is correlated with the Balmer emission.  Similar to the template
fitting in $\S4.1.4$, for each individual spectrum, we mask out the
emission line regions and fit the continuum to a nonnegative linear
combination of a set of SSP templates.  Unlike in $\S4.1.4$, where we
only use models of old and young stellar populations, we here use more
templates, including some of intermediate age as well.  We then
subtract the best-fit model and measure the flux and equivalent width
(EW) of \oiiilam, which is more reliable than the direct \Hb~emission
measurement.  We then correct for the \Hb~emission line assuming
$\Delta \Hb = 0.6~\mathrm {EW}$(\oiiilam) \citep[see,
  e.g.,][]{gonzalez93, trager00a}.  We also correct \Hg~and
\Hd~for emission assuming the decrements $\Hg/\Hb=0.46$ and
$\Hd/\Hb=0.26$.  For reference, the median \Hb{} correction
for all galaxies in the {\tt Elliptical} sample is $\sim
0.37$~\AA, $\sim 4$ Gyr in SSP analysis.

We emphasize that the corrections we adopt are by no
means perfect.  The relation between \Hb~and \oiiilam~ has a large
scatter \citep[see also,][]{trager00a, mehlert00, nelan05,
  kuntschner06}.  For EW(\oiiilam)$=0.5$~\AA, an error of $0.2$ in the
conversion factor translates to $0.1$~\AA~in the $\Hb$ measurement.
On the other hand, the \oiiilam~has been widely used in the literature
and is reliable to measure.

Another way to correct for the emission is to use \Ha~emission
measurements, by adopting the intrinsic Balmer decrement
\Ha/\Hb$=2.86$ and a reasonable dust extinction
\citep[\eg,][]{graves07}.  If dust is well-behaved, this relationship
should be tighter than that for \oiiilam.  However, the \Ha~and
\Hb~emission lines are entangled with the underlying stellar
absorption feature.  In fact, they are almost impossible to measure
reliably if they are relatively weak ($<1$~\AA), which is the case for
most of our spectra.  In particular, because measuring the emission
lines requires modeling the stellar continuum, using corrections based
on \Ha~tends to merely recover the H$\beta$ index of the continuum
template itself.

Here we estimate the emission line infill using \oiiilam~because the
line is easier to measure independently of the stellar continuum, and
because its intrinsic scatter is less important due to the large
sample size in our case.  In any case, we have tried using \Ha~by
adopting E(\Ha-\Hb)$=0.1$ mag and find that the dependence of H$\beta$
absorption-line strength on $\sigma$ and environment is
basically not affected.

\subsubsection{Velocity dispersion dependence of Lick indices}

\begin{figure}
\epsscale{1.10}
\plotone{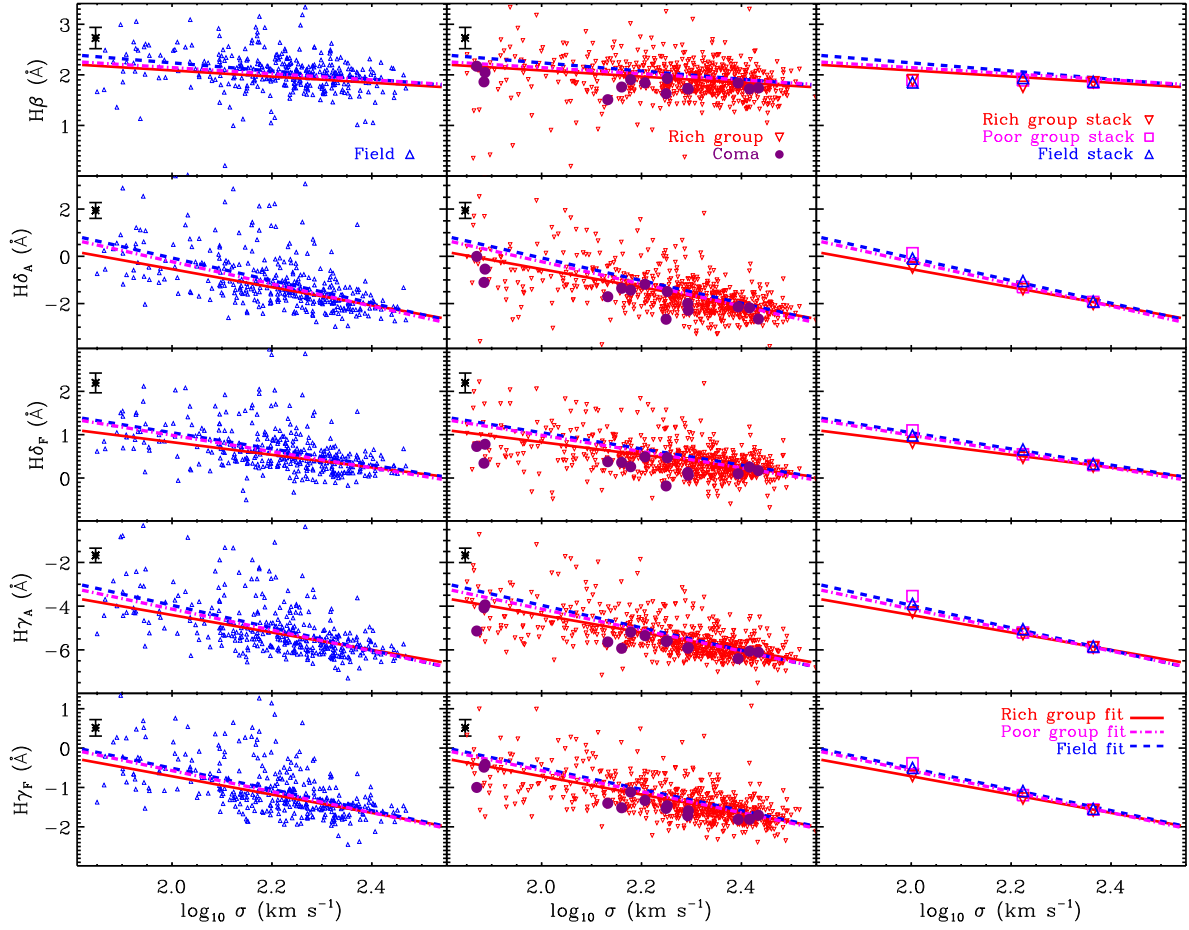}
\caption{
Balmer absorption indices in the Lick indices
as a function of velocity dispersion ($\sigma$) and environment.
We show the measurements for field galaxies with blue open triangles in the left panels,
those for rich-group galaxies with red upside-down triangles in the middle panels and
those for average spectra in the right panels.
We also single out the Coma galaxies as purple solid circles in the middle panels.
The blue dashed lines represent the linear fits to the measurements
for field glaxies, the red solid lines are the linear fits to those for rich-group galaxies,
and the magenta dot-dashed lines are for poor-group galaxies (not shown here).
We show the typical errors of measurements in the upper left corner in each panel.
We correct the emission line infill by assuming EW(\Hb)=$0.6$ EW(\oiiilam) and
higher-order Balmer decrement $\Hg/\Hb=0.46$ and $\Hd/\Hb=0.26$.
The Balmer indices are all weaker at higher $\sigma$
and stronger in the field.
The emission line infill corrections of \Hb~measurements of the average spectra in the lowest $\sigma$ bin appear to
be underestimated (Figure \ref{hbetaoiii}).
}
\label{balmerlick}
\end{figure}

\begin{figure}
\epsscale{1.10}
\plotone{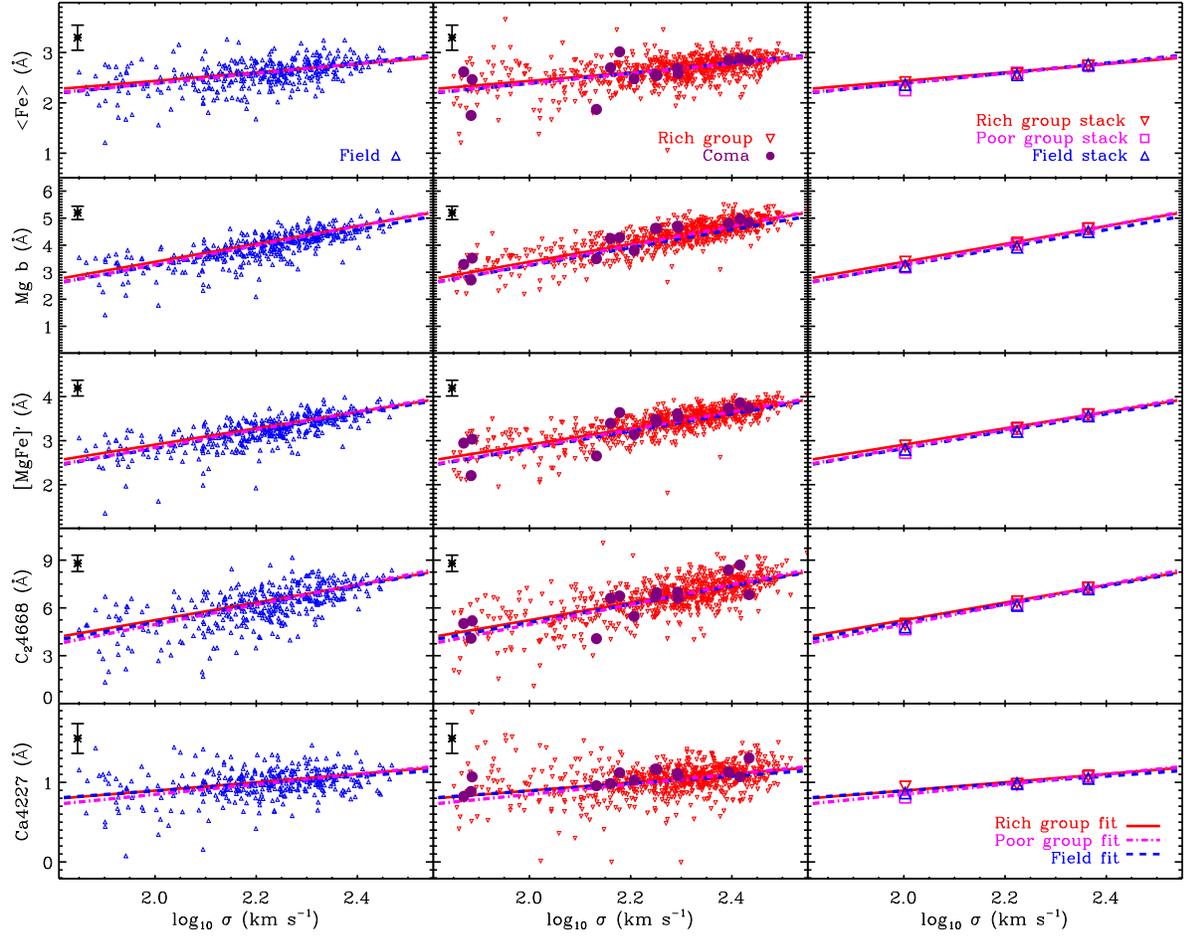}
\caption{
The same as Figure \ref{balmerlick}, but for metallicity indicators in the Lick indices.
The metal absorptions look very similar.
Only the $\alpha$-element indices seem to be slightly weaker in the field, but
at a barely significant level.
}
\label{metallick}
\end{figure}

Figures \ref{balmerlick} and \ref{metallick} present our Lick index
measurements as a function of velocity dispersion and environment.
Figure \ref{balmerlick} shows the Balmer absorption lines: \Hb, \HgA,
\HgF, \HdA, and \HdF.  Figure \ref{metallick} shows the metallicity
indicators: \Fe, the average of \Fefour~and \Fefive, \Mgb, \MgFep,
\Ctwo, and \CAone.  The index \MgFep~is defined as follows: \[
\MgFep=\sqrt{\Mgb\cdot (0.72\cdot {\rm Fe5270}+0.28\cdot {\rm
    Fe5335})} {\rm .}  \] \MgFep~is a good metallicity indicator
almost independent of \aFe~ratio variations (TMB).  
The left columns of these
figures show the field galaxies and the middle columns show the rich-group 
galaxies.  We single out the Coma galaxies in the middle column
and show them separately.  We exclude the poor-group galaxies in these
two columns, though we do show linear fits to their distribution.  We
also show the indices measured on the average spectra in the right
column.

The most obvious feature from these plots is the strong index-$\sigma$
relation, which has been studied intensively in numerous works (e.g.,
\citealt{terlevich81, gorgas90, guzman92, bender93, jorgensen97,
bender98, bernardi98, colless99, jorgensen99, concannon00,
kuntschner00, poggianti01, proctor02, bernardi03b, caldwell03,
worthey03, mehlert03, nelan05}, T05).  We perform a linear
least-squares fit to each distribution with the following form:
\begin{equation}
  \mathrm{Index} = c_1 + c_2~(\log_{10} \sigma - 2.20)\mathrm{,}
\end{equation}
and list the results in Table \ref{licktable}.  We show these fits in
each panel of Figures \ref{balmerlick} and \ref{metallick} for all
three group classifications.

All the Balmer indices strongly anti-correlate with $\sigma$
($c_2<0$).  Their strength is weaker at higher $\sigma$, indicating
older ages or weaker recent star formation for more massive elliptical
galaxies.  Of these indices, \Hb\ is the one most investigators rely on
as the best age indicator, despite the fact that it is the one with
the most significant emission line contamination issues.  In
particular, it is less sensitive to metal lines than are other Balmer
lines \citep[e.g.,][]{korn05}.

The right column of Figure \ref{balmerlick} shows the measurements of
the average spectra.  All such measurements appear to follow the
best-fit scaling relations very well except for the \Hb~measurement in
the lowest $\sigma$ bin.  This discrepancy is possibly due to strong
Balmer emission compared to the \oiiilam, which makes the emission
correction underestimated.

\begin{figure}
\epsscale{1.0}
\plotone{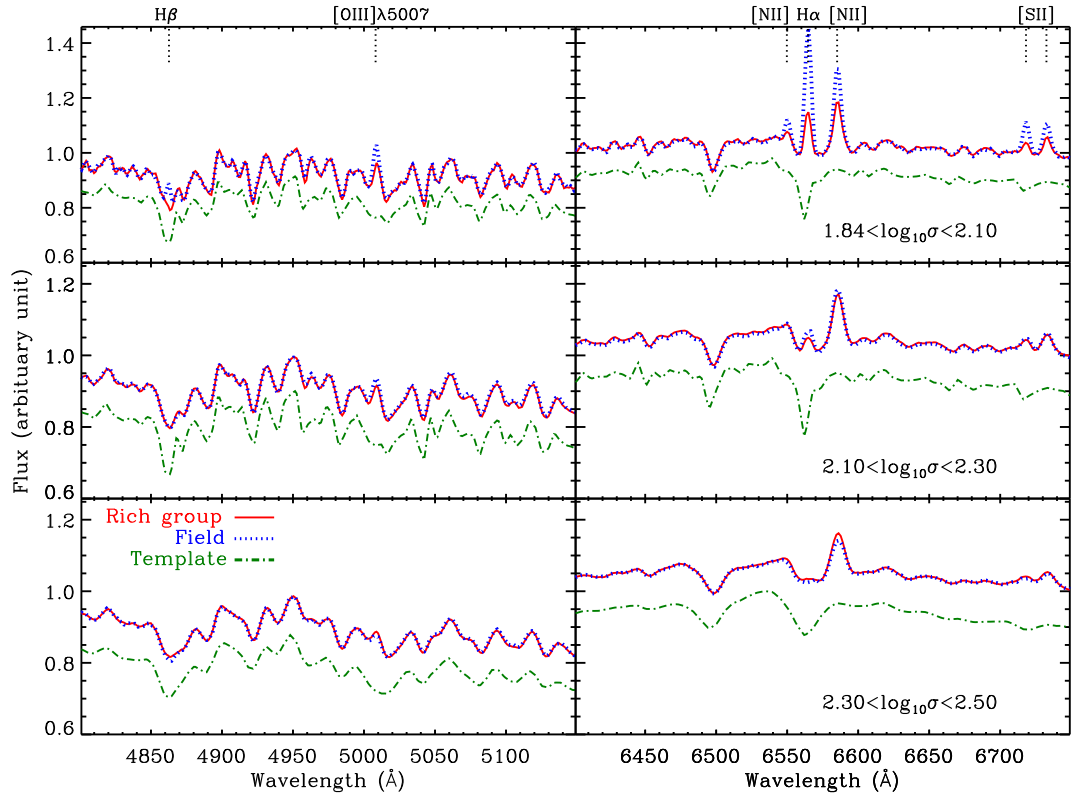}
\caption{
Average spectra within a narrow wavelength range
containing \Hb, \oiii, and \Ha.
Red solid lines represent the average spectra of rich-group galaxies
and blue dotted lines indicate that of field galaxies.
This shows the difficulty in emission line infill correction.
When calculating the Balmer Lick indices, correcting Balmer emission lines 
using direct \Ha~or \Hb~measurements merely recovers the
template used to fit the continuum and it is in fact 
almost impossible to measure reliably
when the emission is weak and entangled with absorption.
Meanwhile, \oiiilam~can be measured more reliably, 
but the \Hb/\oiiilam~ratio suffers from a large scatter.
}
\label{hbetaoiii}
\end{figure}

In Figure \ref{hbetaoiii}, we take a closer look at the average
spectra within the wavelength range containing \Hb, \oiiilam~and
\Ha, where we also overplot the best-fit template to the continuum
for the rich-group galaxies.  The Balmer emission lines are indeed
very strong compared to \oiiilam~in the average spectra of the lowest
$\sigma$ bin, especially for the field galaxies.
Specifically, direct measurements for the field galaxies yield
EW(\Ha)$=2.77$~\AA~and EW(\oiiilam)=$0.87$~\AA.
Using \oiiilam~gives the \Hb~correction $0.52$~\AA, while using
\Ha~gives $0.88$~\AA, larger by $0.36$~\AA.

These average spectra are each stacked from spectra of $\sim100$
galaxies: thus, one galaxy with very peculiar emission lines can
introduce such a feature in the average spectra.  In contrast, the fit
to the scaling relation is less affected by a single data point
outlier, so we will rely below extensively on the scaling relation.
These results emphasize that achieving the emission line
infill correction at the required level of precision is
extremely difficult!

All the metallicity indicators, \Fe, \Mgb, \MgFep, \Ctwo, and \CAone,
correlate with $\sigma$ ($c_2>0$), though to varying degrees.  The
increasing strength with increasing $\sigma$ indicates higher
metallicity for more massive elliptical galaxies.  The tight relation
with $\sigma$ of \Mgb~(slope $c_2 \sim3.44$) is much stronger than
that of \Fe~($c_2\sim0.93$), implying a higher [Mg/Fe] ratio for more
massive galaxies.  Another $\alpha$-element indicator \Ctwo~also
appears to be more strongly correlated with $\sigma$ than \Fe~with a
slope $c_2 \sim5.78$. 
Taken together, these results are evidence of stronger \Ae~in
more massive elliptical galaxies.  In fact, T05 show that increasing
metallicity, \Ae~ and older age account for the Mg-$\sigma$ relation
by $60\%$, $23\%$ and $17\%$ respectively \citep[see
  also][]{mehlert03}.
The dependence of \CAone~on $\sigma$ appears to be much weaker than
other $\alpha$-elements, with a slope $c_2 \sim0.58$, which implies 
that more massive galaxies may be more calcium-underabundant 
\citep[see, \eg][]{thomas03b}.

To summarize, we see signs that the stellar populations in more
massive elliptical galaxies are older or with weaker recent star
formation, more metal-rich, and more strongly $\alpha-$enhanced, as
has been found in the past.  We will come back to this in the next
sections when we interpret our index measurements using SSP models.

\subsubsection{Environmental dependence of Lick
  indices}\label{sec:envdep} 

Figure \ref{balmerlick} shows that the Balmer index-$\sigma$ relation
for elliptical galaxies varies as a function of environment.  The Balmer
lines are systematically stronger ($\sim0.1$~\AA~for \Hb) in field
galaxies than in rich-group galaxies.  The difference vanishes at
high $\sigma$.  This agrees well with the environmental dependence
we see in the continua of the average spectra (Figures
\ref{avgspecratio} and \ref{fieldtogroup}).  We also find that the 
Coma galaxies in our sample all have weaker Balmer lines than average.

Balmer indices are thought to be the best age indicators because
massive stars (\eg, A stars) in a young stellar population
exhibit strong Balmer-line absorption in their integrated
spectra.  The environmental dependence thus suggests that the stellar
populations in field galaxies are in general younger than those in
group galaxies.

In contrast, the distributions of metallicity indicators are
remarkably similar for elliptical galaxies in different environments.
The \Fe~index is nearly identical for all the samples.  The
$\alpha$-element indices seem to be slightly stronger in rich-group
galaxies, but only at a barely detectable level (as seen previously
by, e.g., \citealt{bernardi98}).  However, the same index measurement
does not necessarily mean the same metallicity.  At a given $\sigma$,
if group galaxies are older than field galaxies as the Balmer indices
indicate, then having the same metallicity index strength means they
are more metal-poor.

In this section, we studied the Lick indices, their relation with
$\sigma$, and their environmental dependence.  Consistent with
previous studies, we have found strong index-$\sigma$ relations,
indicating higher metallicity, stronger \Ae~and older ages for more
massive elliptical galaxies.  We have also found that Balmer indices
are systematically weaker in group galaxies, implying older ages for
group galaxies.  We do not see an obvious dependence on environment
for \Fe.  The $\alpha$-element indices (\Mgb, \Ctwo~and \CAone) show
slightly stronger absorption in group galaxies, but only at a barely
detectable level.  This result suggests that group galaxies are
slightly more iron-poor and slightly more strongly
$\alpha-$enhanced than field galaxies.

However, to convert the measurements to more quantitative parameters
of the stellar populations introduces the uncertainties associated
with the stellar population synthesis.  Later in the paper, we will
nonetheless make an attempt to use state-of-the-art SSP models to
study the stellar populations in elliptical galaxies.
 
\subsubsection{Systematic Uncertainties}

We conclude this subsection by summarizing the systematic
uncertainties affecting our Lick index measurements.

We measure the Lick indices on the flux-calibrated spectra that are
smoothed to $\sigma=325~\kms$ under the SDSS resolution $69~\kms$.  We
have used empirical corrections to correct the indices back to the
effective Lick IDS resolution.  However, the corrections may have large
uncertainties \citep{trager98}.  The standard Lick indices are
measured on spectra that are not flux-calibrated.  Therefore, any
direct comparison of our measurements with standard ones should be
taken with caution.

The SDSS spectra are taken with fibers that enclose an aperture with a
diameter of $3''$.  Since we limited our sample to be lower than
redshift $0.05$, the spectra are of the central region of the
galaxies.  The velocity dispersion is therefore essentially the
velocity dispersion in the central region.  For galaxies of different
angular sizes, the aperture therefore encloses the central regions of
different sizes.  Recent work has shown that the radial profile of
$\sigma$ follows a power law $\sigma_{R} \propto {R}^{\gamma}$ where
$R$ is radius and the index $\gamma \sim -0.04$ to $-0.07$
\citep[e.g.,][]{jorgensen95, mehlert03, cappellari06}.  The effective
radii of the {\tt Elliptical} sample have a median value $\sim 7''$.  If we
correct all the $\sigma$ to that at effective half light radius with
such a scaling relation, this translates to a difference of $\sim3\%$;
this small effect should not affect the analysis presented here.

The Lick indices measured here also come from the spectra of stellar
populations in the central regions, with a much smaller fraction from
those in the outer region within the aperture.  The Lick indices also
probably correlate with radius, and the gradients of indices such as
\Mgb~and \Fe~ seem to correlate with that of $\sigma$ as well.  On the
other hand, \Hb~is roughly constant with radius
\citep[e.g.,][]{mehlert03, kuntschner06, sanchez06c, rawle10}.  The
correlations are still poorly understood and we do not correct
for them.

Finally, the emission line infill correction for the Balmer indices of
an individual spectrum is extremely difficult, even in the case of
stacked spectra with $\sim 100$ galaxies --- one galaxy with a
peculiar emission line ratio can introduce strong errors in the
correction.  Nevertheless, we are using \oiiilam~to correct Balmer
emission lines, which should be a good compromise for such a large
sample.  We also have tried using \Ha~too, which basically does not
affect the relative dependence on $\sigma$ and environment. 

All of these factors may introduce some errors at the few percent
level in our measurements.  We do not expect they introduce any
significant systematic bias in the differential aspect of our
analysis, because they should not correlate strongly with which
subsample we consider.

\subsection{Comparison of Lick indices with SSP models}

\subsubsection{SSP models}

One purpose of stellar population synthesis models is to place
constraints on the history of star formation and chemical enrichment
of galaxies from their integrated SEDs.  We apply two state-of-the-art
versions of these models to our measurements: those of TMB and
S07.

There are a few caveats that make our goal very challenging.  For
example, age and metallicity have extremely similar effects on the
color and SEDs
\citep[\eg,][]{faber72, faber73, oconnell80, rose85, renzini86,
worthey94}; Lick indices have very broadly-defined line windows which
makes direct translation into element abundances difficult
\citep[\eg,][]{greggio97, tantalo98, korn05, serven05}; finally, many
elliptical galaxies have non-solar [\aFe] which further complicates the
modeling \citep[\eg,][]{peletier89, worthey92, davies93, mcwilliam94}.

Recent SSP models have incorporated adjustable abundance patterns for
multiple elements and allow for more reliable derivation of age,
metallicity, and element ratio from absorption-line indices (\eg,
\citealt{borges95, weiss95, tantalo98, trager00b}, TMB,
\citealt{maraston05, coelho07}, S07).  
Most of these models, however, are based on the Lick system
and are not directly applicable to our data, as discussed in
\S\ref{sec:licksmoothed}.  
Nonetheless, S07 provide empirical zero-point corrections to correct
flux-calibrated Lick indices onto the standard Lick system (see their
Table 1).  We apply these corrections to our Lick index measurements
when we compare them with the TMB models.  The S07 models, meanwhile,
are built up on flux-calibrated spectral library, so no corrections
are necessary for that comparison.

The corrections necessary for the comparison to TMB have a very
large scatter (see Table 1 and Figure 1 of S07).  Therefore, while we
will compare to TMB below, our main focus will be on the models of S07.

\subsubsection{Comparison with TMB models and S07 models}

Before deriving the stellar population parameters of our
sample with SSP models, we first compare our measurements to the
predictions of each model on grids of age, metallicity and
$\alpha-$enhancement.

One of the main differences between TMB models and S07 models is that
TMB models are built at fixed metallicity [Z/H], while S07 models are
at fixed iron abundance [Fe/H].  To convert between [Z/H] and [Fe/H],
we adopt the relation given by TMB: [Z/H]=[F/H]+$0.94$\,[\aFe].
Because we tie other $\alpha$-elements to [Mg/Fe] when building up S07 
models (see next subsection), we assume [Mg/Fe]=[\aFe]~in the comparison.

\begin{figure}
\epsscale{1.0}
\plotone{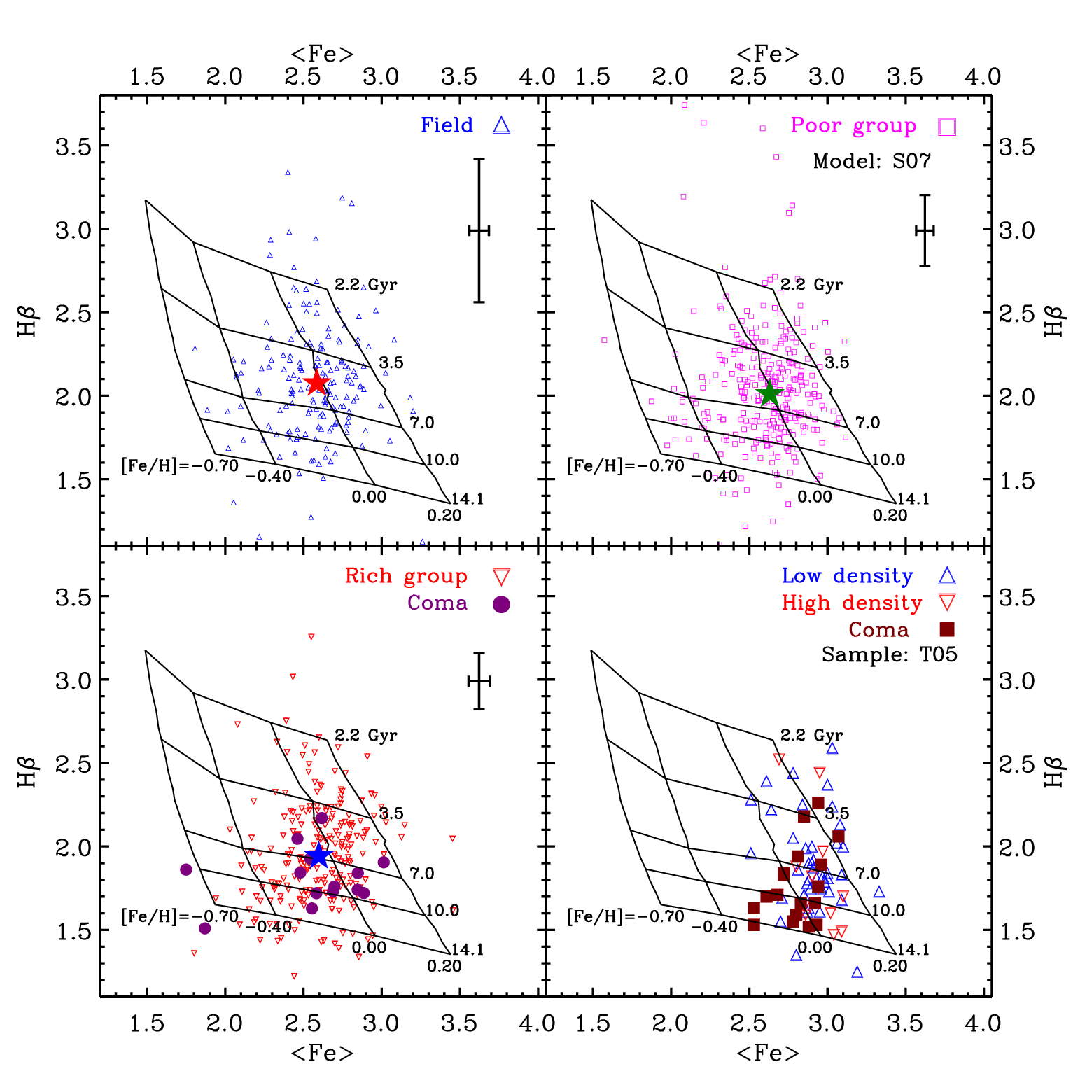}
\caption{
\Hb~ vs. \Fe~diagram.
We plot \Hb~and \Fe~on top of the S07 models with [Mg/Fe]=$0.30$.
We here only show the galaxies within
the velocity dispersion range $2.10<\log_{10}\sigma<2.30$ ($125~\kms<\sigma<200~\kms$).
The filled stars are the median values of field galaxies, poor-group galaxies and 
rich-group galaxies in each panel. The error bars show the $\osigma$ scatters in both axes.
In the lower right panel, we also show the elliptical galaxies in the T05 sample,
corrected to flux-calibrated Lick indices with empirical corrections given by S07.
Compared to rich-group galaxies, field galaxies are more spread into younger grids.
The Coma galaxies in our sample appear to be in general older than average.
Our \Hb~measurements are systematically larger than those in T05 sample, most likely caused
by emission correction.
}
\label{hbetafe}
\end{figure}
\begin{figure}
\epsscale{1.0}
\plotone{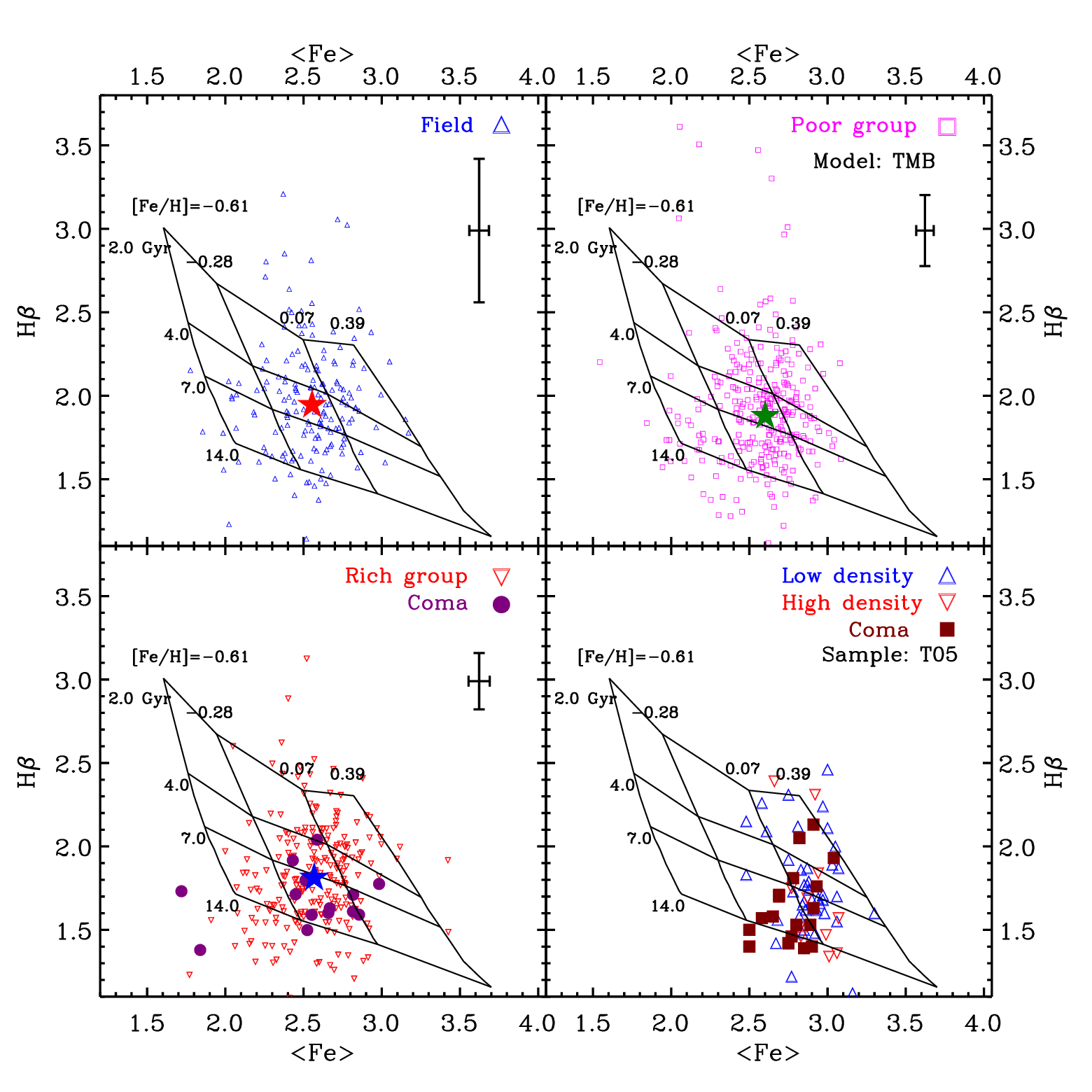}
\caption{
\Hb~ vs. \Fe~diagram, but with TMB models with [\aFe]=$0.30$.
We correct all the measurements to the standard Lick indices with empirical 
corrections given by S07 to match the TMB models.
In the lower right panel, we also show the elliptical galaxies in the T05 sample.
}
\label{hbetafetmb}
\end{figure}

In Figure \ref{hbetafe} and \ref{hbetafetmb}, we compare our measurements 
of \Hb~versus \Fe~with the S07 and TMB models assuming
[\aFe]=[Mg/Fe]=$0.3$.
We choose [$\alpha$/Fe]=$0.3$ because most 
elliptical galaxies have super-solar [$\alpha$/Fe].  
In the following analysis we 
only use elliptical galaxies with $\sigma$ between $125~\kms$ 
and $200~\kms$ ($2.10<\log_{10}\sigma<2.35$, the middle $\sigma$ bin),
to ensure that we do not confuse $\sigma$ dependence with environmental
dependence.  The median $\sigma$ values within this range are
$166~\kms$, $167~\kms$ and $172~\kms$ for field, poor-group, and rich-group 
galaxies, respectively. We also show the median measurements of
each subsample as filled stars and the $\osigma$ scatters as 
error bars below the legend.  Finally, for comparison,
in the lower right panel of each figure we compare the physical
parameters of the galaxies in our sample to those of the early-type
galaxy sample of T05.

In general, the S07 and TMB models agree well (see S07;
\citealt{graves08, kuntschner10a} for more thorough comparisons
between the two models).  We see that the galaxies in our {\tt
  Elliptical} sample have a median SSP-equivalent age of $\sim 7$ Gyr.
These measurements show that group galaxies, especially rich-group
galaxies, appear to occupy a different locus than field galaxies.
In particular, the distribution of group galaxies extends to older
ages and lower iron abundance: relative to the other samples, the 
rich-group galaxies appear to more commonly have sub-solar iron abundance
([Fe/H]$<0$) and older age ($\gtrsim 7.0$ Gyr); meanwhile, the field
galaxies are typically $\lesssim 7.0$ Gyr.  This result
suggests that the rich-group galaxies are slightly older and slightly
more iron-poor (in terms of [Fe/H]).  We also show the T05 sample in
the bottom right panel and see very similar effects.  Note when
comparing with S07 models, we have corrected measurements of the T05
sample to the flux-calibrated indices.  We also see that the Coma
galaxies are systematically older than average.

\begin{figure}
\epsscale{1.0}
\plotone{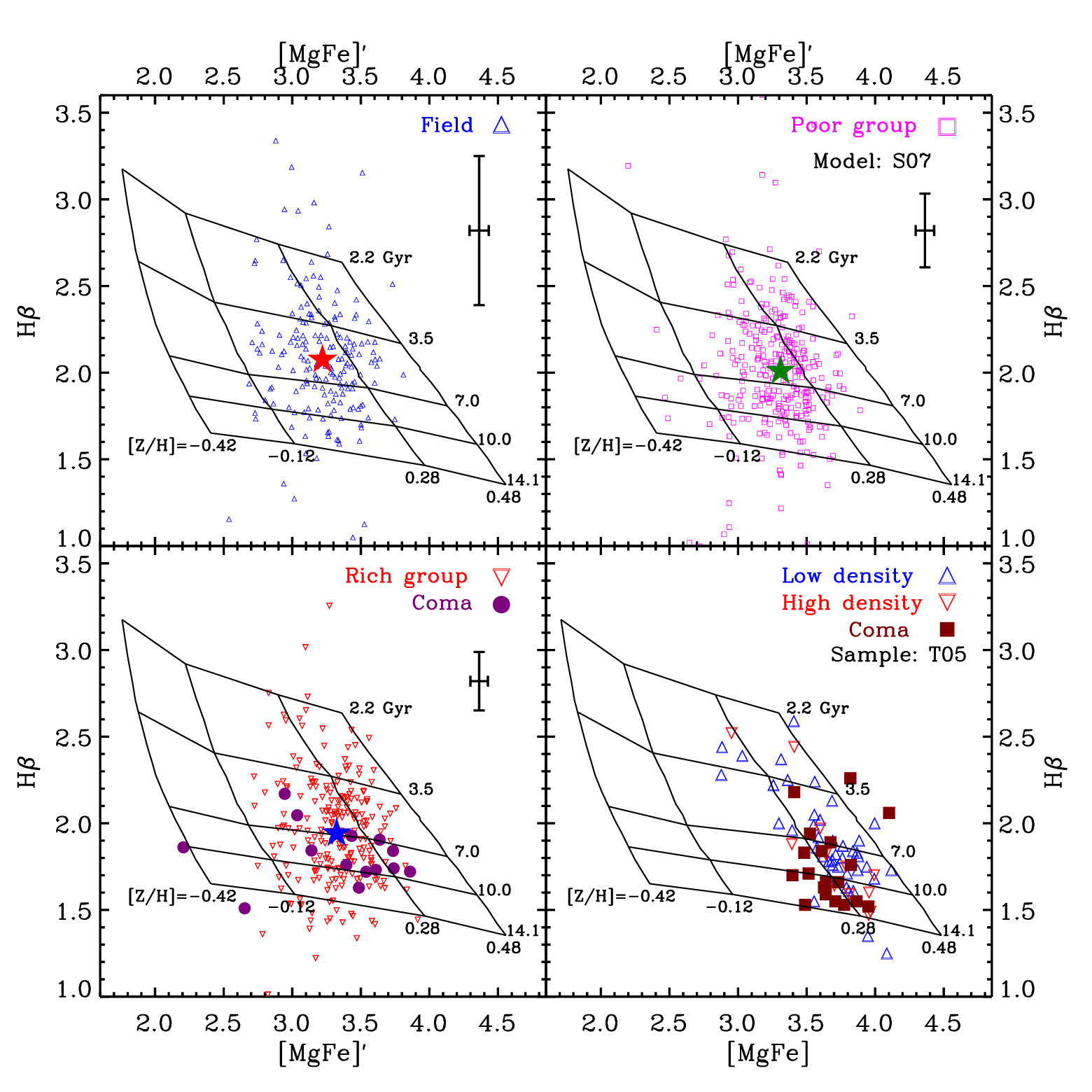}
\caption{
\Hb~ vs. \MgFep~diagram.
We plot the measured \Hb~and \MgFep~on top of the S07 models with 
[Mg/Fe]=$0.30$.  We only show the galaxies within
the velocity dispersion range $2.10<\log_{10}\sigma<2.30$ ($125~\kms<\sigma<200~\kms$).
The filled stars are the median values of field galaxies, poor-group galaxies 
and rich-group galaxies in each panel.The error bars show the $\osigma$ scatters in both axes.
In the lower right panel, we also show the elliptical galaxies in the T05 sample,
corrected to flux-calibrated Lick indices with empirical corrections given by S07.
Because the T05 catalog only includes the average index \Fe, we instead show the
index \MgFe=$\sqrt{\Mgb \cdot \Fe}$.
}
\label{hbetamgfe}
\end{figure}

\begin{figure}
\epsscale{1.0}
\plotone{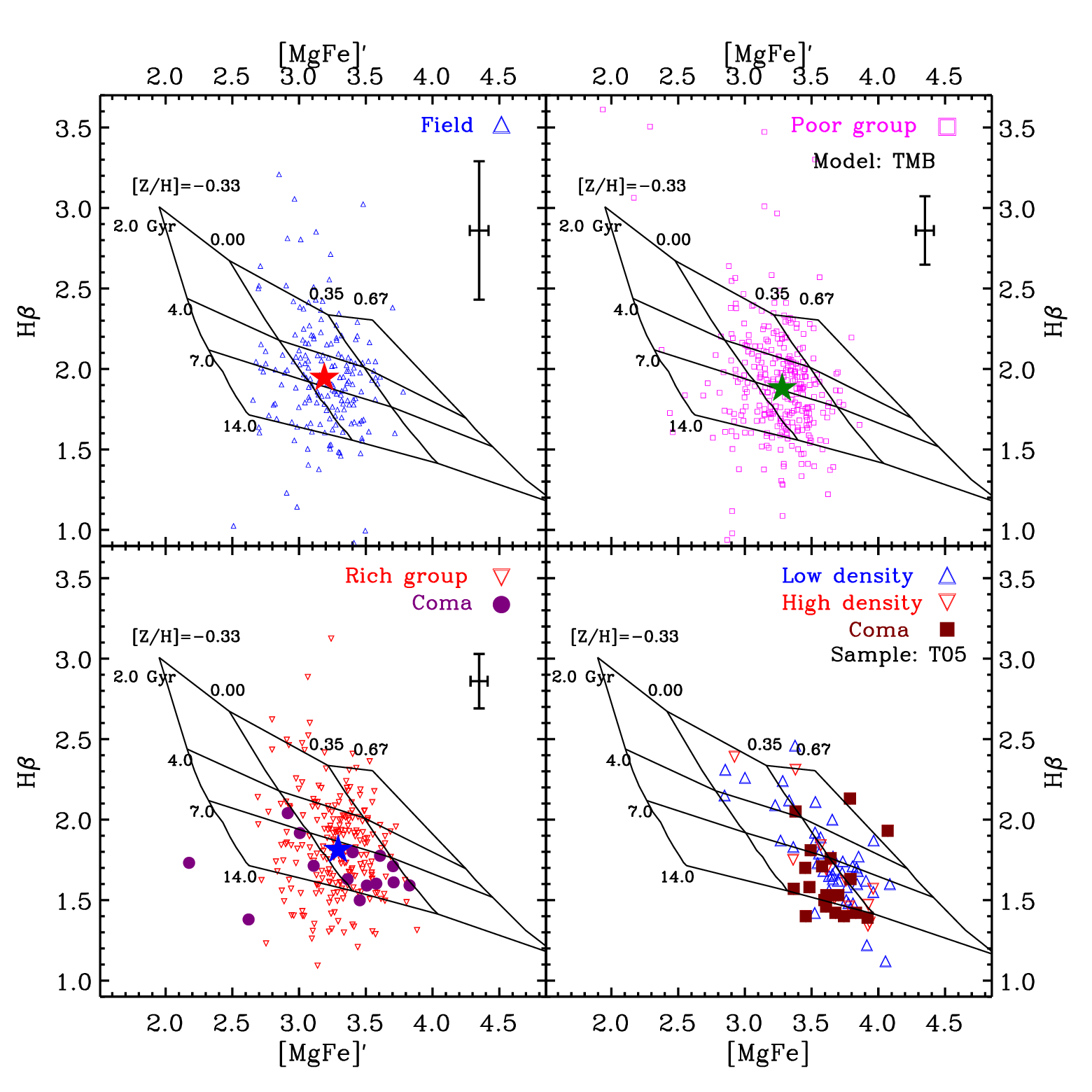}
\caption{
\Hb~ vs. \MgFep~diagram, but with TMB models with [\aFe]=$0.30$.
We correct all the measurements to the standard Lick indices with empirical 
corrections given by S07 to match the TMB models.
In the lower right panel, we also show the elliptical galaxies in the T05 sample.
}
\label{hbetamgfetmb}
\end{figure}

The strength of the \Fe~index is mainly determined by the iron
abundance.  For total metallicity, \MgFep~is a better indicator.  In
Figure \ref{hbetamgfe} and \ref{hbetamgfetmb}, we plot our
measurements \Hb~vs. \MgFep~against S07 and TMB models with
[$\alpha$/Fe]=$0.3$.  These measurements show that in different
environments the total metallicities [Z/H] are very similar.  The T05
sample\footnote{The public catalog of T05 only includes the average
index \Fe, so we show instead the combination \MgFe=$\sqrt{\Mgb
\cdot \Fe}$.  However, the difference is very small.}, however,
exhibit slightly lower
total metallicity for ellipticals in high-density environments.

\begin{figure}
\epsscale{1.0}
\plotone{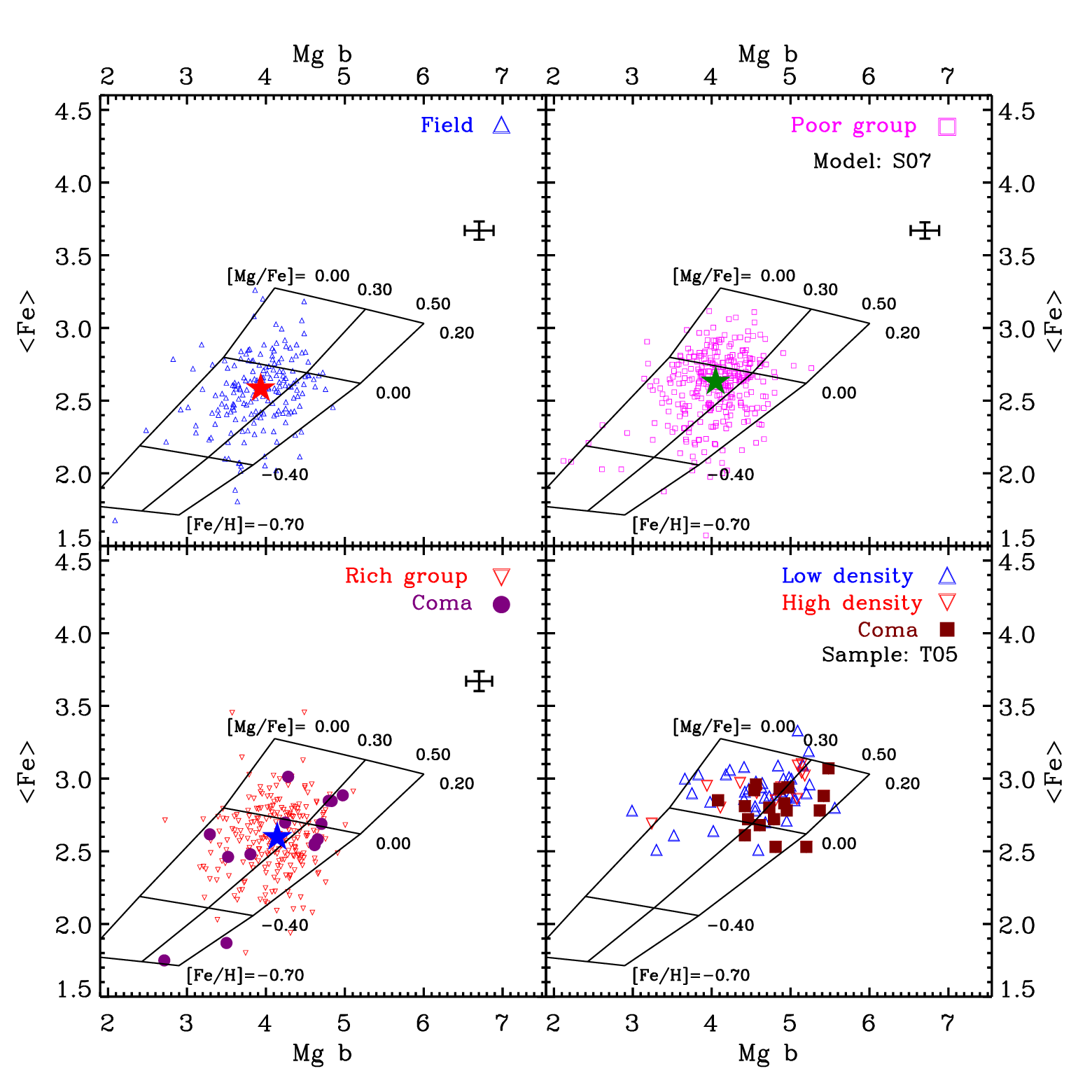}
\caption{
\Fe~vs. \Mgb~diagram.
We plot the measured \Fe~and \Mgb~on top of the S07 models with age $7$ Gyr.
We only show the galaxies within
the velocity dispersion range $2.10<\log_{10}\sigma<2.30$ ($125~\kms<\sigma<200~\kms$).
The filled stars are the median values of field galaxies, 
poor-group galaxies and rich-group galaxies in each panel. 
The error bars show the 1-$\sigma$ scatter in both axes.
In the lower right panel, we also show the elliptical galaxies in the T05 sample,
corrected to flux-calibrated Lick indices with empirical corrections given by S07.
Rich-group galaxies are slightly more spread into higher [Mg/Fe] grids.
}
\label{femgb}
\end{figure}

\begin{figure}
\epsscale{1.0}
\plotone{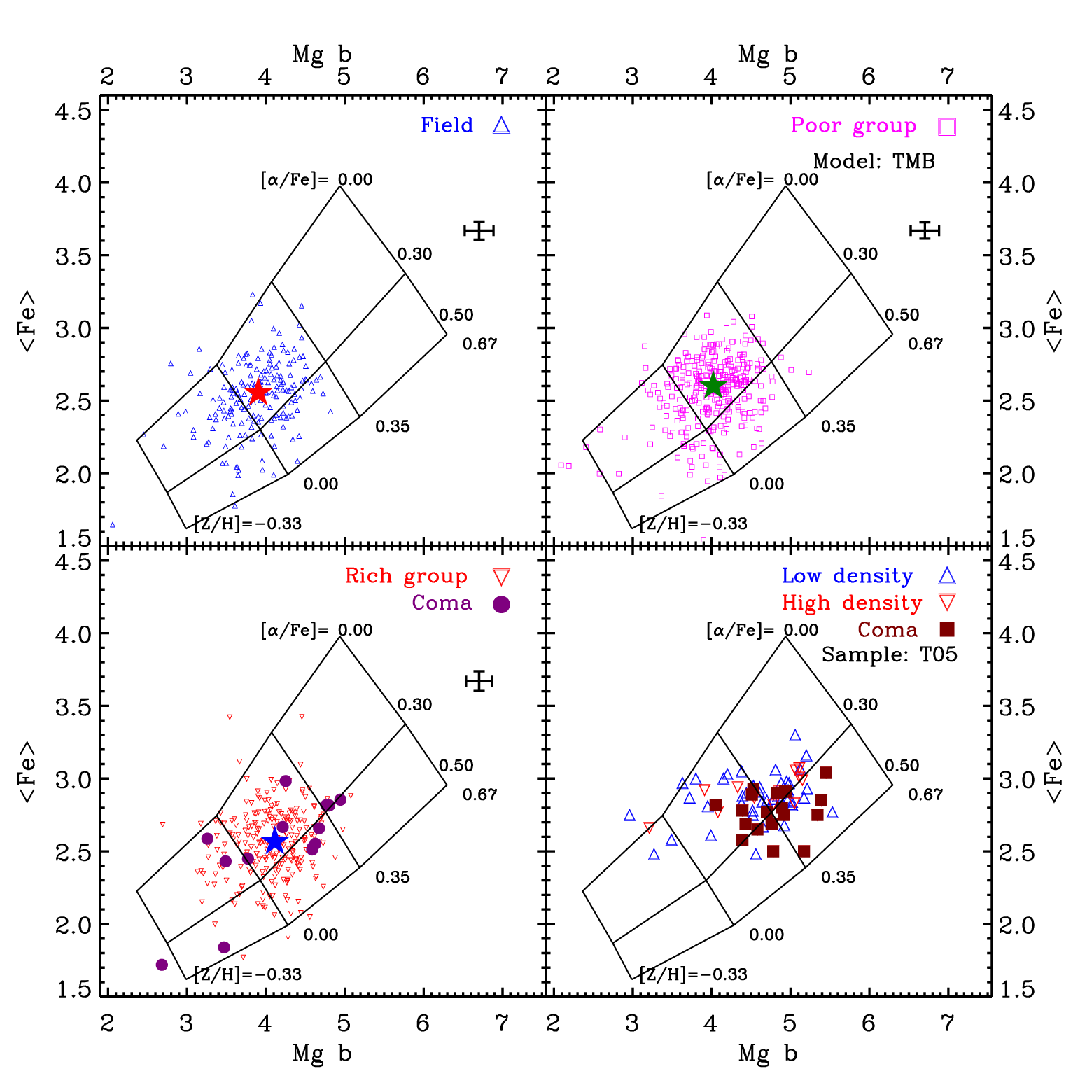}
\caption{
\Fe~vs. \Mgb~diagram, but with TMB models with age $7$ Gyr.
We correct all the measurements to the standard Lick indices with empirical 
corrections given by S07 to match the TMB models.
In the lower right panel, we also show the elliptical galaxies in the T05 sample.
}
\label{femgbtmb}
\end{figure}

In Figures \ref{femgb} and \ref{femgbtmb}, we compare our measurements
of \Fe~and \Mgb~against S07 and TMB models with an age of $7$
Gyr.  We see that almost all elliptical galaxies have super-solar
[Mg/Fe], with a median value close to $0.3$, and the
elliptical galaxies in rich groups exhibiting somewhat higher
[Mg/Fe].  In the lower right panel, we also show the elliptical
galaxies in the T05 sample.  Although T05 do not claim to see an
environmental dependence in their analysis, which includes lenticular
galaxies (not shown here), we see the elliptical galaxies in
high-density environments in their sample also appear to be slightly
more strongly $\alpha-$enhanced.

\subsubsection{SSP-equivalent parameters: age, [Fe/H], [Mg/Fe] and [Z/H]}

Here, we use S07 models to derive the SSP-equivalent parameters.  To do
so, we create a grid of models and interpolate our results onto the
grid.

First we use the {\tt deltabund} code in the {\tt EZ\_ages} package to
create SSP models on a grid of three parameters: age, [Fe/H] and
[Mg/Fe].  We use the solar-scaled isochrones as suggested in the {\tt
EZ\_ages} documentation, and the \citet{salpeter55} IMF.  We set [O/Fe]
to be zero for these isochrones and tie other $\alpha$-elements to Mg.
The age allowed in the models ranges from $1.2$ Gyr to $17.7$ Gyr, the
[Fe/H] ranges from $-1.3$ to $0.2$ and we generate models at [Mg/Fe]
between $-0.1$ and $0.5$.  We refer the readers to S07 and
\citet{graves08} for details about S07 models and the {\tt EZ\_ages}
package.
 
Once we have all the models, we employ a two-step interpolation method
to derive age, [Fe/H] and [Mg/Fe].  For each galaxy, we first
calculate age and [Fe/H] using \Hb~and \Fe, at all [Mg/Fe] (from -0.1
to 0.5).  We then compare the pair \Fe~and \Mgb~with models at the
median age we obtained in the first step to calculate a new [Fe/H] and
[Mg/Fe].  We then update the age by interpolating the ages found in
the first step at the [Mg/Fe] we found in the second step and iterate
the second step with the new age.  The iteration usually only needs
$\sim 2$ steps before fully converging.  After convergence, we have the
parameters age, [Fe/H] and [Mg/Fe] for each galaxy with
\Hb, \Fe~and \Mgb.  We also give [Z/H] assuming [Z/H]=[F/H]+$0.94$\,[\aFe].
For those measurements that fall out of the 
model grids, we set the parameter to be the boundaries, i.e., the
maximum or the minimum in the models.  
We have also tried using higher-order Balmer lines instead of \Hb~and 
found very consistent results.

\begin{figure}
\epsscale{1.0}
\plotone{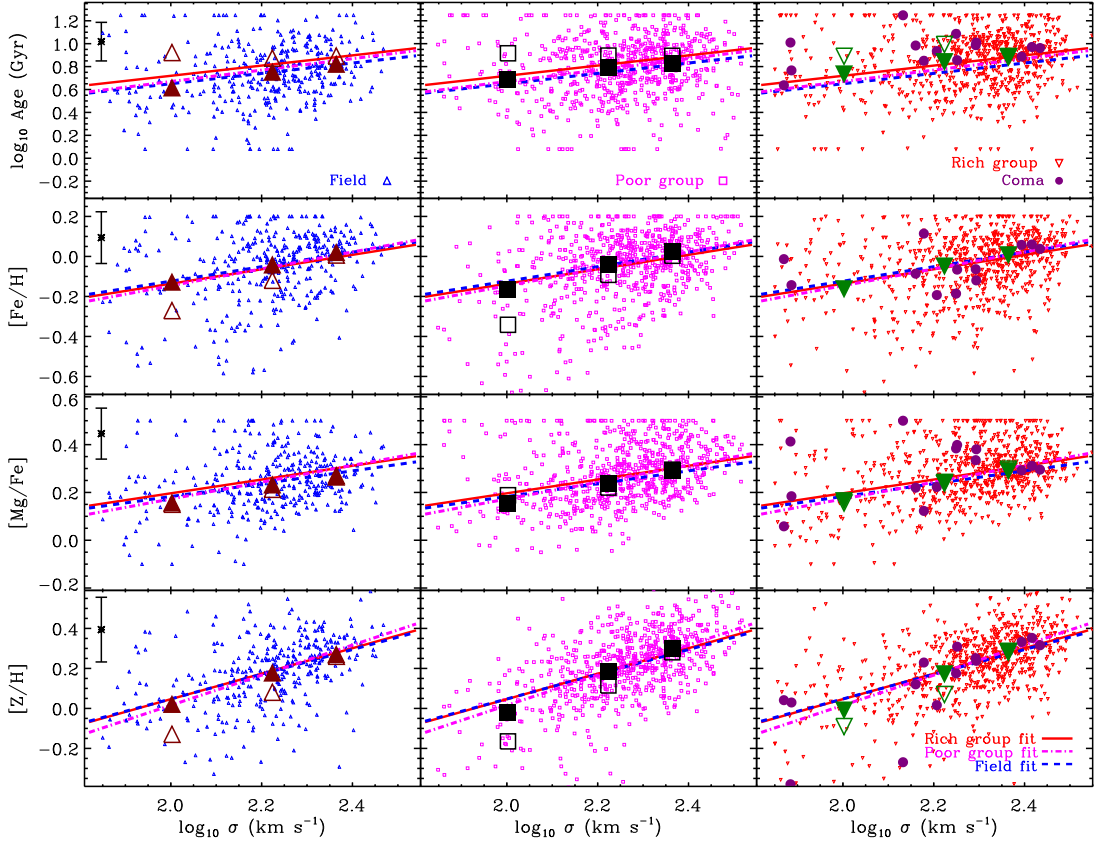}
\caption{
\label{sspvsenv}
SSP-equivalent parameters vs. velocity dispersion ($\sigma$).
Using \Hb, \Fe~and \Mgb, we derive the SSP-equivalent parameters from the S07 models with 
age, [Fe/H], [Mg/Fe] and [Z/H]=[Fe/H]+$0.94$\,[\aFe].
The red solid lines are the linear fits to the derived parameters for rich-group galaxies,
the magenta dot-dashed lines represent those for poor-group galaxies and
the blue dashed lines are for field galaxies.
The large open symbols show the derived parameters for the average spectra,
whose derived age and [Fe/H] are
affected by the uncertainties in emission line infill correction of \Hb.
The large filled symbols represent the derived parameters
at the same velocity dispersion of the average spectra,
but assuming the index-$\sigma$ scaling relations (Eq. 1) hold exactly.
The error bars in the upper left corner in the left panels are the median
Monte Carlo errors.
All parameters strongly correlate with $\sigma$, but all with a large scatter.
More massive galaxies are older, more metal-rich and more strongly $\alpha-$enhanced.
Rich-group galaxies are systematicall older than field galaxies, by $\sim1$ Gyr,
although this effect is most pronounced at low $\sigma$.
Rich-group galaxies also appear to be slightly more iron-poor (in terms of [Fe/H])
and slightly more strongly $\alpha-$enhanced, but only at a barely detectable level.
There is no noticeable difference of total metallicity [Z/H] in different environments.
}
\label{ssppar}
\end{figure}

Figure \ref{ssppar} shows all the measurements as a function of
environment and $\sigma$.  We also overplot the derived parameters for
the average spectra as large open symbols.  In addition, we also show
as large filled symbols the derived parameters at the same $\sigma$ of
the average spectra but assuming that the index-$\sigma$ relations
(Eq. 1 and Table 2) hold exactly.

As we pointed out above, the \Hb~absorption in the average spectra is
underestimated due to the underestimation of the emission correction.
The derived age therefore is overestimated, and the derived [Fe/H]
underestimated.  For this reason, in this context we trust results
based on the index-$\sigma$ relationships more than the average
spectra.

The most obvious feature in Figure \ref{ssppar} is the strong
dependence on $\sigma$ of all three parameters, as we expect from the
index-$\sigma$ relations (though with a large scatter due to the
combined errors from the three indices).  More massive elliptical
galaxies are older, more metal-rich and more strongly
$\alpha-$enhanced.  The environmental effect on age is also apparent:
galaxies in groups appear to be older than their counterparts in the
field.  Meanwhile, we do not see an obvious environmental effect on
[Fe/H], [Mg/Fe] and [Z/H], implying the environmental dependence of [Fe/H],
[Mg/Fe] and [Z/H] is very subtle, if it exists.

To quantify the $\sigma$ and environmental dependence of the
SSP-equivalent parameters we fit the scaling relation 
for each distribution as a
function of $\sigma$ as follows:
 \begin{equation}
\label{ssplinfit}
  \mathrm{SSP~parameters} = s_1 + s_2~(\log_{10} \sigma - 2.20)\mathrm{.}
 \end{equation}
However, we need error measurements for the derived parameters in the
fitting.  The errors are unfortunately more difficult to quantify.

We assume all the measurement errors are independent and
propagate the errors to the derived parameters with a Monte Carlo
method.  To do this, we create $50$ fake measurements for each galaxy
by adding Gaussian errors to the measurements of \Hb, \Fe~and \Mgb,
and derive the new parameters for the fake measurements.  We then
calculate the standard deviations of the parameters of the Monte Carlo
ensemble for each galaxy and quote it as the Monte Carlo error of the
derived parameters.

We show the median Monte Carlo error of each parameter in the upper
left corner in Figure \ref{ssppar}.  However, the individual error of
each point can be heavily biased, due to edge effects on the grid.
Almost all the galaxies have some simulated data points falling
outside the grid.  This effect is more severe for those galaxies
with parameters closer to the boundaries.  In the extreme case, the
simulated measurements of a galaxy with index measurements lying
outside the grid are almost always also outside the grid; thus, for
the Monte Carlo ensemble of such a galaxy, the parameters derived with
our method are almost always the same as the boundary values in the
models.  Therefore, the Monte Carlo errors are unrealistically small for
the parameters of the galaxies with measurements close to or outside
the model boundaries.

However, for most of the derived parameters that are relatively far
away from the boundaries, the Monte Carlo errors are very similar.  We
therefore apply a uniform error (the median Monte Carlo error) to each
parameter of all galaxies.  We also ignore the galaxies with
measurements that are out of grids when fitting the scaling relation,
because they have the same values for the derived parameters (either
the maximum or the minimum) at all $\sigma$, which will heavily skew
the fitting.

Using these choices, we fit equation (\ref{ssplinfit}) to the derived
parameters.  Figure \ref{ssppar} shows the best-fit scaling relations
and Table \ref{ssptable} lists all of the coefficients.  Besides the
errors in the slopes and intercepts, we have also calculated the
internal scatter of the distribution of the whole sample in our three
$\sigma$ bins.  The internal scatter (in dex) for $\log_{10}{\rm age}$, [Fe/H]
and [Mg/Fe] in the three $\sigma$ bins (in ascending order) are: 
($0.22$, $0.21$, $0.18$), ($0.24$, $0.14$, $0.10$)
and ($0.13$, $0.11$, $0.08$), respectively.
Recall that the median $\sigma$ in the three bins are $100$, $167$ and
$232~\kms$, respectively.

The strong $\sigma$ dependence is obvious from the best-fit relations.
Elliptical galaxies at the highest masses are older, more metal-rich,
and more strongly $\alpha-$enhanced.  The rich-group galaxies are also
apparently older than the field galaxies, by $\sim1$ Gyr.  We
note, however, that the difference between different subsamples is
relatively small compared with the intrinsic scatter.  In the
previous sections, we showed that the environmental dependence of the
average spectra and
\Hb-$\sigma$ relations vanishes at higher $\sigma$.  We therefore
expect that the SSP-equivalent age dependence on environment may
correlate with $\sigma$ too.  Indeed, the slope of the age-$\sigma$
relation for field galaxies is slightly larger than that for 
rich-group galaxies, but the difference is very small.  The dependence of
the age-environment relationship on $\sigma$ thus may be buried in the
large scatter in this diagram.

Aside from the strong $\sigma$ dependence and the obvious
environmental dependence of age in the best-fit scaling relations, we
do see very subtle environmental effects on [Fe/H] and [Mg/Fe].
However, these are at a barely significant level ($\osigma$).
And we do not see obvious dependence of total metallicity [Z/H] on environment.
Compared to the field galaxies, the rich-group galaxies are slightly more
iron-poor (only by $\sim0.01$ dex in terms of [Fe/H]) and slightly
more strongly $\alpha-$enhanced (only by $\sim0.01$ dex).  These
differences are consistent with the environmental dependence of the
index-$\sigma$ relations we saw in \S\ref{sec:envdep}.  In
Figure~\ref{metallick} and Table~\ref{licktable}, we show that the
\Fe~absorption-line strength is almost identical in different
environments.  However, because field galaxies are younger,
they need slightly more iron to produce the same \Fe~signature as
the rich-group galaxies; field galaxies also have slightly weaker
\Mgb~(and other $\alpha$-element indices), implying slightly lower
$\alpha-$abundance in the field, as we see here.

The environmental dependence we see agrees very well with
previous studies (\eg, T05; \citealt{bernardi06, cooper10}).  T05 show
that early-type (E/S0) galaxies in high-density environments (mostly
in the Virgo and the Coma clusters) are older by $\sim2$ Gyr than
galaxies in low-density environments.  We have also shown that the
Coma galaxies in our {\tt Elliptical} sample are systematically older
than average; thus, the Coma galaxies in their sample may have
contributed to the relatively stronger dependence in their results.
They also found a relatively stronger dependence for metallicity; in
particular, they found that galaxies in high-density environments are
more metal-poor by $~0.05-0.1$ dex (in terms of [Z/H]) with no
dependence for [$\alpha$/Fe].  As seen in Figure \ref{femgb}, if we
only look at the elliptical galaxies in their sample, we see the
galaxies in high-density environments appear to be slightly more
strongly $\alpha-$enhanced.  Considering the large scatter in the SSP
parameter-$\sigma$ relation, our results agree very well with each
other.

By comparing samples in different environments, \citet{bernardi06}
find that the spectroscopic differences between early-type galaxies in
high-density environments and in low-density environments are very
similar to the differences between early-types at redshift $z=0.17$
and $z=0.06$.  Under the reasonable assumption that the primary
difference between the samples at each redshift is overall age, they
therefore conclude that early-type galaxies in high-density
environments are $\sim 1$ Gyr older than those in low-density
regions.  Thus, our conclusions are in very good agreement with theirs.
They also do not find noticeable environmental dependence in total metallicity.
Combining results from different groups (\eg, T05, \citealt{bernardi06}, and
this work), we note that the environmental dependence in metallicity is indeed very
subtle.

Using samples drawn from the SDSS, \citet{cooper10} remove the 
mean dependence of average overdensity (i.e., environment) on color and luminosity,
and find that there remains a strong residual trend between
stellar age and environment, such that galaxies with older stellar populations 
favor regions of higher overdensity relative to galaxies of like color and 
luminosity. Their conclusions therefore are also consistent with the environmental 
dependence we find here.

\citet{thomas10} do not find a noticeable environmental
dependence for the bulk old population of early-type
galaxies, likely because they separate the young
objects with signs of recent star formation from the bulk old population. 
In \S\ref{sec:losfr} we argued that the environmental dependence of 
the average spectra was likely due to recent low-level star formation 
in low $\sigma$ galaxies.  We conclude, therefore, that our results 
would be consistent with \citet{thomas10} if they had included the young
early-type galaxies in their scaling-relation analysis.

\begin{figure}
\epsscale{1.0}
\plotone{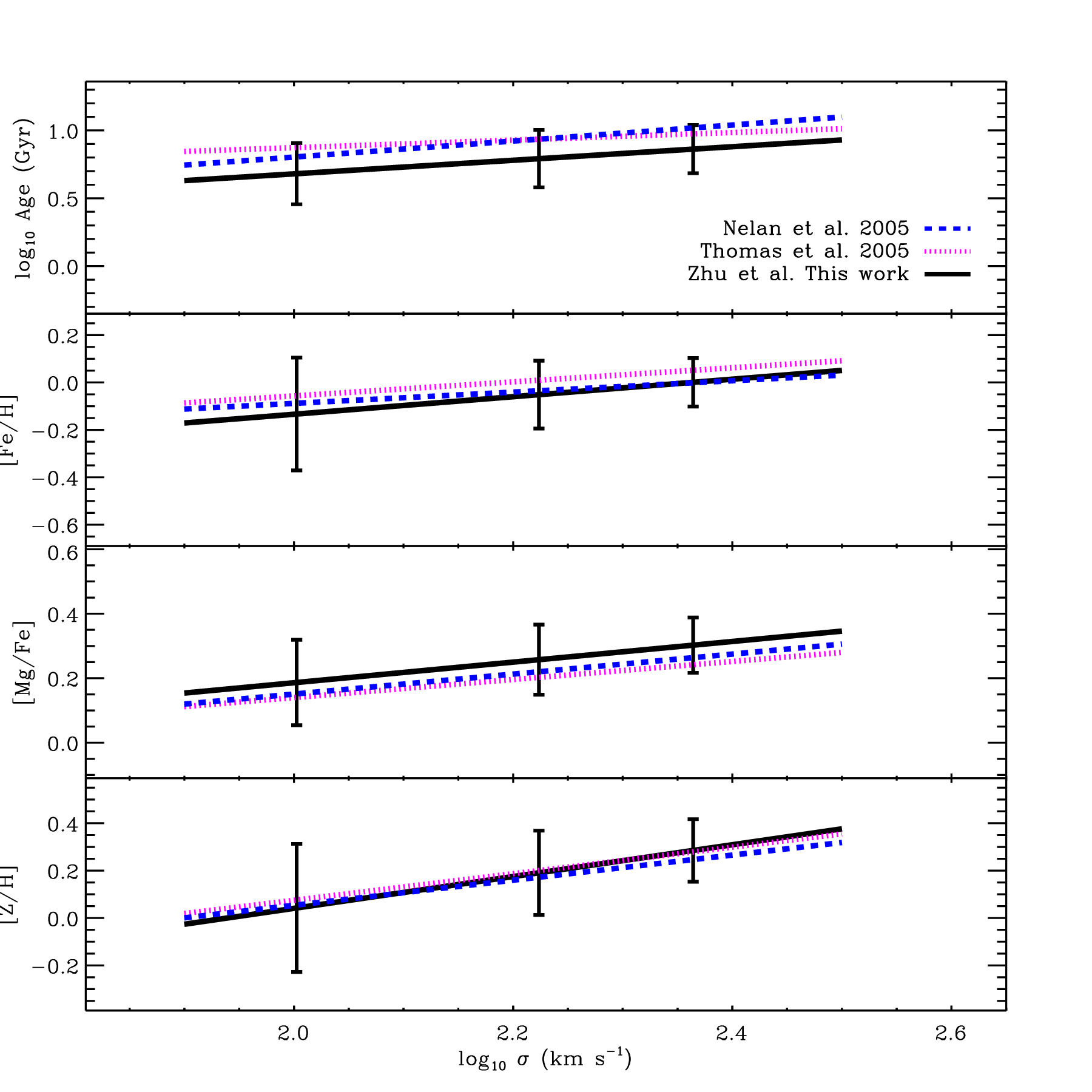}
\caption{
Comparison of the best-fit scaling relations between
SSP-equivalent parameters and velocity dispersion ($\sigma$) with those derived in \citet{nelan05} and T05.
The error bars are the $1\sigma$ scatter for each parameter in the whole sample, not
the errors of the fits in Table \ref{ssptable}.
We have assumed [$\alpha$/Fe]=[Mg/Fe] and [Z/H]=[Fe/H]+$0.94$\,[\aFe] (TMB)
in the comparison.
The relations are in good agreement with each other.
The main difference is in the age.  Our derived age is $\sim3$ Gyr younger than theirs,
mainly because our \Hb~measurements are larger than
theirs by $\sim0.3$ \AA~(see, e.g., Figure \ref{hbetafe}).
The difference is mostly likely caused by emission line infill correction (see text).
The median emission correction for \Hb~in our sample is $0.37$ \AA, which translates to $\sim 4$ Gyr in the derived age.
}
\label{sspparcomp}
\end{figure}

In Figure \ref{sspparcomp}, we compare the linear relationships
we derive between velocity dispersion and age, iron abundance,
$\alpha$-enhancement, and total metallicity against those derived by
T05 and \citet{nelan05}.
We emphasize that the fits derived by both T05 and
\citet{nelan05} include elliptical and lenticular galaxies, 
while our sample only includes elliptical galaxies.  
We determine the intercept of \citet{nelan05} from their Table $8$ and
Figure $13$.  For T05, we use the mean of the slopes and the
intercepts for samples in high-density and low-density environments.
We also overplot the $\osigma$ scatter in the distribution of the
parameters for the whole sample as error bars (though of course the
errors in the mean values are much smaller).

Figure \ref{sspparcomp} shows that our slopes are in very good
agreement with theirs.  Considering the large scatter in the
distribution, uncertainties in zero-point corrections, different
samples and measurements used, and different models used, we find the
agreement is remarkable and very encouraging for SSP analysis.  The
main difference is that our age-$\sigma$ relation is
systematically younger, by $\sim3$ Gyr.  This is mainly caused by the
difference in the \Hb~measurements in that our \Hb~indices are
systematically stronger than theirs, by $\sim0.3$~\AA~(see, \eg,
Figure \ref{hbetafe}).

The discrepancy in age can probably be attributed to the emission line
infill correction and/or sample bias.  We notice that in
\citet{nelan05} they exclude galaxies with significant emission lines,
by which they may have thrown away galaxies with significant young
stellar populations.  For the resulting sample without significant
emission lines 
, they do not correct for emission-line contamination.  In the
sample compiled by T05, we find in the literature that the indices of
some galaxies are corrected for emission \citep{gonzalez93}, while
others are not (\citealt{beuing02}, \citealt{mehlert00}\footnote{We
  are not sure whether or not they correct emission for the galaxies
  with strong emission (EW(\Hb)$>0.3$~\AA).}).  Note the median
correction for \Hb~in our {\tt Elliptical} sample is $\sim0.37$~\AA.
Although this difference is fairly small compared to the absorption
strength ($\sim2$~\AA), it translates to $\sim 4$ Gyr in SSP analysis.
In addition, in T05 more than half of the galaxies in high-density
environments are from the Coma cluster.  As we have shown here, the
Coma galaxies in our sample are systematically older than average.
Taking into account the emission line infill correction, sample bias,
and the large scatter of the distribution, we conclude that these
age$-\sigma$ relationships agree fairly well with each other.

 \subsubsection{Caveats on SSP parameters}

We emphasize that the parameters calculated above are SSP-equivalent
parameters.  That is, they can only be taken literally if elliptical
galaxies formed in a short enough, and uniform enough, burst of star
formation.  Such a scenario is probably incorrect.  Recent
observations from GALEX, {\it Spitzer}, and HST have shown that a
significant fraction of early-type galaxies exhibit strong UV excess,
PAH emission and IR excess, implying possible low-level recent star
formation \citep{yi05, rich05, kaviraj07b, schawinski07, temi09,
  young09, salim10}.  Similarly, in \S\ref{sec:losfr}, we showed that the
environmental dependence of our frosting model parameters (\eg,
\citealt{trager00b, gebhardt03}, S07) were consistent with the results
of \citet{schawinski07} who used GALEX NUV photometry to look for
recent star formation in early-type galaxies.

If recent star formation is responsible for some or all of the
$\sigma$ and environmental dependence of the average spectra, then
instead of giving the age of the dominant old stellar population, the
derived SSP-equivalent age will be significantly younger (than the
real age of the old population) because of the existence of a
significant fraction of young stars \citep[\eg][]{trager00b, serra07}.
\citet{serra07} show that the Balmer indices of a composite stellar population
depends primarily on the mass fraction and age of the younger
component.  If so, then the $\sigma$ and environmental dependence of
the derived age we see here may reflect partly, or totally, the
$\sigma$ and environmental dependence of the mass
fraction and age of the young component in the frosting model.  That
is, elliptical galaxies at different $\sigma$ may host the old base
stellar populations of the same (or similar) age, but the
SSP-equivalent ages are younger at lower $\sigma$ and in the field 
due to stronger recent star formation.  

Meanwhile, because of the age-metallicity degeneracy, the derived
metallicity might be overestimated due to the underestimated age ---
assuming the metallicity indices are primarily tracing the old
population.  Precisely how the metal absorption indices behave in the
case of composite stellar populations is as yet unclear.  

\section{Discussion}

In the above sections, we have found systematic differences in the
average spectra of ellipticals, as a function of $\sigma$ and
environment.  In addition, we have measured the dependence of
SSP-equivalent ages, [Fe/H], [Mg/Fe] and [Z/H] on these parameters.

As previous studies have shown, stellar ages of elliptical galaxy seem to be
older at higher masses.  In this sense, our results fit into the
now-classic ``downsizing'' scenario that \citet{cowie96a} presented
in the context of the more general galaxy population.  This age
dependence on $\sigma$ appears to be explainable in the hierarchical
scenario of galaxy formation, as long as star formation can be shut
down effectively (e.g., \citealt{delucia06}).  In these models, the
higher mass halos are comprised of small systems that on average
collapsed earlier.  If there is a mechanism to shut off their
star formation quickly enough, these progenitors become gas poor and
thus the higher mass systems will have older stellar population ages.

The precise nature of what shuts off star formation in elliptical
galaxies is unknown.  AGN are a possible source for such feedback,
through both radiative and mechanical heating processes
(\citealt{ciotti97a,benson03a, delucia06, vernaleo06a, ciotti09a}).  A
commonly invoked mechanism for triggering the AGN is by feeding the
central black holes during a major merger (e.g., \citealt{silk98a,
dimatteo05a, hopkins08a}).  Indeed, \citet{hopkins08a} argue that in
these models the time since the last major gas-rich merger is an
increasing function of galaxy mass, and could explain an age-$\sigma$
relationship of ellipticals.  Other mechanisms for shutting off
star formation also are potentially important: \citet{ciotti09a} fuel
AGN feedback using the recycled gas from dying stars, while
\citet{johansson09a} and others invoke gravitational heating from
infalling stellar systems. 

Elliptical galaxies also have a strong dependence of
$\alpha-$abundance on $\sigma$.  This trend implies that the nature of
chemical enrichment depends on galaxy mass.  The iron-peak elements
mainly come from Type Ia supernovae, while the $\alpha$-elements are
mainly produced in massive stars experiencing Type II supernovae.
That the $\alpha-$enhancement (or probably more accurately, the
iron-deficit) is stronger in more massive elliptical galaxies may
imply that their star forming timescale is shorter than less massive
elliptical galaxies, before the delayed Type Ia supernovae enrich the
star forming regions with iron-peak elements 
(\eg,\citealt{thomas98, pipino04}; T05. But also see
\citealt{smith09a}).  Hierarchical models with feedback, such as those
of \citet{delucia06}, appear consistent with this result at least
qualitatively.

Finally, ellipticals have a dependence of metallicity on stellar mass.
If most of the stars are formed in situ, this trend might be explained
by their deeper potential well, boosting chemical recycling by
hampering the outflow of gas from the galaxy
\citep[\eg,][]{arimoto87, matteucci94}.  However, if most of the
stellar mass in ellipticals is produced in progenitor spiral disks,
this metallicity trend may instead be a remnant of the
mass-metallicity trends known for spirals (e.g. \citealt{garnett02a,
pilyugin04a, tremonti04a}).

Where the stars formed that now constitute elliptical galaxies remains
in debate.  \citet{naab09a} claim that the metallicities of present day
spirals are several times lower than those measured for ellipticals,
indicating that giant ellipticals cannot have been produced from
mergers of spirals.  \citet{hopkins09b} counter that this problem could
be mitigated if the stellar metallicity gradients of ellipticals are
large enough; typical metallicity measurements of ellipticals only
cover apertures that are a fraction of the effective radius, and the central
regions might have highly enhanced metallicities.  It remains to be
determined what would be the mass-metallicity relation predicted by a
fully cosmological simulation of the merger scenario (though see
\citealt{kobayashi05a, finlator08a}).

A final wrinkle in the dependence of elliptical properties as a
function of mass is that their strength and scatter put some
constraint on their late-time merger history.  After all, if the
largest galaxies are built from progenitor populations of lower-$\sigma$
galaxies, one expects the dependence of age and
chemical abundance on $\sigma$ to weaken with time.  If the merging
process is important to galaxy growth, these trends should be 
steeper and tighter in the past.  An interesting quantitative prediction from
the theoretical models would be how large this evolution really was
expected to be, given various merging scenarios.

Alternatively, a variation of the stellar initial mass function (IMF) with
galaxy mass --- within the ellipticals or their progenitors --- can
account for the metallicity and $\alpha-$abundance trends as well
(\citealt{cenarro03a, nagashima05a, koppen07a, hoversten08a}).  

Our results demonstrate that whatever processes produce these
correlations differ across environment, though only slightly.
Explaining the differences between ellipticals in rich groups and in
the field requires either a slightly later time of formation for field
galaxies, or equivalently and more likely a slightly thicker frosting
of recent star formation.  Furthermore, these differences are
substantially stronger for low-mass galaxies than for high-mass
galaxies.  This distinction between the effects on high- and low-mass
ellipticals is apparent both in the stellar continuum shapes, and in
the age measurements based on the Lick index scaling relations (the
age measurements we trust the most; large filled symbols in Figure
\ref{sspvsenv}).

As noted above, the stellar ages of ellipticals may be related to the
formation times of their progenitor halos.
\citet{gao05a} \citep[see also][]{wechsler06a, zhu06}
show that CDM dark matter halos in dense environments
were assembled earlier than average.  Such an effect may also be
reflected in Fig.~1 of \citet{delucia06}, of the stellar age as a
function of environment.  If the ages of the populations are related to
the formation time of the progenitor halos, these theoretical results
could explain the younger populations seen in the field.  The
theoretical results also predict that this effect declines at higher
masses (reversing, in fact, above the current nonlinear mass).  These
results could be related to the environmental dependence of mean
stellar age.

Alternatively, field ellipticals may simply have somewhat better
access to reservoirs of cold inflowing gas whose virial temperature in
groups would be too hot to accrete at late times.  In this case, the
mean stellar age difference would derive mainly from recent
star formation, which as we note above is actually seen in nearby
fast-rotator ellipticals \citep[\eg,][]{kaviraj07b}.  However, it is
unclear whether and why such an effect should be weaker for more
massive ellipticals, as required by the data.  For example, the
gravitational heating scenario of \citet{khochfar08a} appears to
predict a fairly strong dependence of age on environment for massive
galaxies.  These considerations possibly favor internal processes,
rather than external ones, for regulating the star formation rates in
the most massive ellipticals.

The metallicity and $\alpha$-abundance of stellar populations are a
much weaker function of environment than age.  Whatever determines the
chemical evolution of such galaxies must therefore be a weak function
of environment.  For example, if the mass-metallicity relation is
determined by a systematic variation in the IMF, 
our results would imply that the IMF does not vary substantially 
with environment.
In contrast, if merger history is important in
establishing the present day mass-metallicity relation, then
ellipticals in different environments must have similar enough merger
histories. On the other hand, if the environmental dependence of
age is caused by recent star formation (\S\ref{sec:losfr}), we expect
that stronger recent star formation would cause lower [\aFe]~and higher 
metallicity in field ellipticals.
For example, by separating the young early-types from the bulk 
old population, \citet{thomas10} find that the young population
have lower [\aFe]~and higher metallicity.  It is possible that our 
sample is still too small to detect a stronger environmental dependence. 
It is therefore of great interest to investigate the relation of 
metallicity and $\alpha$-abundance with environment with a larger 
sample of elliptical galaxies.

\section{Summary}

We reanalyzed the images of red-sequence galaxies in the local
universe using the homogeneous data set of the SDSS.  By carefully
examining the surface brightness profiles, we selected $1,923$
elliptical galaxies with velocity dispersion $\sigma>70~\kms$ 
at redshift $z<0.05$.  We found that
elliptical galaxies dominate the bright/massive end ($\gtrsim L^\ast$)
of the red sequence in the color-magnitude diagram, and disk-dominated
galaxies dominate at lower luminosity ($\lesssim L^\ast$).  We also
found that elliptical galaxies and disk-dominated galaxies form different loci
in color gradient-magnitude/velocity dispersion space, suggesting
color gradient can be used for morphological classification.

We studied the dependence of properties of elliptical galaxies on
velocity dispersion and environment.  Group galaxies tend
to have higher velocity dispersion, and thus higher mass.

We have calculated the average optical spectra as a function of
velocity dispersion and environment, which all show a typical SED of
an old stellar population.  We found that the average spectra depend
strongly on velocity dispersion.  Elliptical galaxies at lower-$\sigma$
have a bluer continuum, stronger Balmer and
nebular emission, and weaker metal absorption.  This result is
consistent with the color-magnitude/mass relation that brighter/more
massive elliptical galaxies are redder in optical broad-band colors.

Interestingly, we found weak but significant environmental dependence
of the average spectra.  Elliptical galaxies in the field have a bluer
continuum, especially at wavelengths $\lesssim 4000$~\AA, and have
stronger (but still weak) emission lines than their counterparts in
groups.  However, this dependence on environment appears
primarily for low-$\sigma$ ellipticals; the highest-$\sigma$
ellipticals are much less affected.  Assuming
that age is the dominant factor shaping SEDs and that elliptical
galaxies consist of an old base stellar population and a small
frosting of young stars, we fit the frosting model to each spectrum in
the {\tt Elliptical} sample.  Assuming these trends to be due to a
young population, we found that the fraction of galaxies with
significant young population is higher at lower velocity dispersion
and in the field. We also found that the environmental dependence 
appears primarily at low velocity dispersion and vanishes at high 
velocity dispersion.

We measured the Lick indices of the flux-calibrated SDSS spectra of
galaxies in our {\tt Elliptical} sample.  In agreement with previous work,
we found strong index-$\sigma$ relations.  The Balmer absorption
indices are stronger at lower velocity dispersion, while the
metallicity indices are stronger at higher velocity dispersion.  The
Balmer indices of field galaxies are systematically stronger than
those of rich-group galaxies.  As in the case of average spectra, 
this dependence is most pronounced at low velocity dispersion and disappears
at high velocity dispersion. We did not find significant
environmental dependence of metallicity indicators, only that the
$\alpha$-element indices appear to be slightly stronger in rich-group
galaxies.  We also noted that the emission line infill correction for
Balmer lines is an extremely difficult task.

Assuming SSP analysis applies to elliptical galaxies, we have derived
the SSP-equivalent age, iron abundance (in terms of [Fe/H]), 
$\alpha-$enhancement ([Mg/Fe]) and total metallicity ([Z/H]).  
We found that the SSP-equivalent
parameters strongly correlate with velocity dispersion.  More massive
elliptical galaxies are older, more metal-rich, and more strongly
$\alpha-$enhanced.  We also found that, galaxies in rich groups are
systematically older than their counterparts in the field, by $\sim1$
Gyr. However, this effect is strongest at low velocity dispersion.
We found that galaxies in rich groups are slightly more
iron-poor and slightly more strongly $\alpha-$enhanced, but only at a
barely significant level.  And we do not find noticeable environmental dependence
of total metallicity as well. We have performed fits to the
parameter-$\sigma$ relation and found that our fits are in very good
agreement with previous work, especially taking into account many
uncertainties such as zero-point corrections, different samples and
data, and different SSP models employed.  We found that emission line
infill corrections can affect the age determination by $\sim 3-4$
Gyr.  We also caution that the SSP-equivalent age and metallicity may
be affected by the existence of the young component if the frosting
model applies to elliptical galaxies.

Finally, we make our sample publicly available to the
community\footnote{\tt
http://bias.cosmo.fas.nyu.edu/galevolution/elliptical}.

\acknowledgments

We are grateful to Ricardo Schiavon and Genevieve J. Graves for making
their code and models publicly available and for very useful discussions.  
We wish to thank Daniel Thomas for making their models and sample publicly available. 
We also thank him for various comments that helped improve the
paper at the refereeing stage.
We would also like to thank Philip Hopkins, Thorsten Naab, and Jeremiah Ostriker for
useful discussions.

The authors acknowledge funding
support from NSF grant AST-0607701, NASA grants 06-GALEX06-0030,
NNX09AC85G and NNX09AC95G, and \emph{Spitzer} grant G05-AR-50443.
 
This research has made use of NASA’s Astrophysics Data System and of the NASA/IPAC 
Extragalactic Database (NED) which is operated by the Jet Propulsion Laboratory, California 
Institute of Technology, under contract with the National Aeronautics and Space Administration. 
    
 Funding for the SDSS and SDSS-II has been provided by
 the Alfred P. Sloan Foundation,
 the Participating Institutions,
 the National Science Foundation,
 the U.S. Department of Energy,
 the National Aeronautics and Space Administration,
 the Japanese Monbukagakusho,
 the Max Planck Society,
 and the Higher Education Funding Council for England.
 The SDSS Web Site is http://www.sdss.org/.
 
 The SDSS is managed by
 the Astrophysical Research Consortium for the Participating Institutions.
 The Participating Institutions are
 the American Museum of Natural History,
 Astrophysical Institute Potsdam,
 University of Basel,
 University of Cambridge,
 Case Western Reserve University,
 University of Chicago,
 Drexel University,
 Fermilab,
 the Institute for Advanced Study,
 the Japan Participation Group,
 Johns Hopkins University,
 the Joint Institute for Nuclear Astrophysics,
 the Kavli Institute for Particle Astrophysics and Cosmology,
 the Korean Scientist Group,
 the Chinese Academy of Sciences (LAMOST),
 Los Alamos National Laboratory, 
 the Max-Planck-Institute for Astronomy (MPIA),
 the Max-Planck-Institute for Astrophysics (MPA),
 New Mexico State University,
 Ohio State University, University of Pittsburgh,
 University of Portsmouth,
 Princeton University,
 the United States Naval Observatory,
 and the University of Washington.
 

\newpage
\clearpage
\begin{deluxetable}{ccc}
\tabletypesize{\small}
\tablecolumns{5}
\tablewidth{0pc}
\tablecaption{Sample Definition}
\tablehead{
\colhead{Sample}  & \colhead{Size} & \colhead{Description} }
\startdata
Lowz & $87,623$ & All galaxies with SDSS imaging below z $< 0.05$ \\ 
Environ & $57,885$ & Galaxies with $M_r < -19.0$ in Lowz \\
PhotoRS & $37,026$ & Galaxies that pass red sequence cuts ($\S 2.1$) \\
SpecRS & $32,726$ & Galaxies with SDSS spectroscopy in PhotoRS-Sam. $22,621$ with $\sigma>70~\kms$ \\
Elliptical & $1923$ & The elliptical galaxy sample \\
Bonus & $430$ & The bonus elliptical galaxy sample without SDSS spectroscopy\\
\enddata
\label{sampletable}
\end{deluxetable}

\newpage
\clearpage
\begin{deluxetable}{cccccc}
\tabletypesize{\scriptsize}
\tablecolumns{5}
\tablecaption{Index-$\sigma$ relation}
\tablehead{
\colhead{Index} & \colhead{Coeffecient} & \colhead{All} & \colhead{Rich group} & \colhead{Poor group} & \colhead{Field}}
\startdata
 \Hb~(\AA)      & $c_1$ & $ 2.02\pm 0.01$ & $ 1.97\pm 0.01$ & $ 2.02\pm 0.01$ & $ 2.08\pm 0.01$ \\
 \Hb~(\AA)      & $c_2$ & $-0.71\pm 0.04$ & $-0.61\pm 0.06$ & $-0.61\pm 0.07$ & $-0.81\pm 0.10$ \\
 & & & & & \\
 \HdA~(\AA)     & $c_1$ & $-1.18\pm 0.01$ & $-1.31\pm 0.02$ & $-1.17\pm 0.01$ & $-1.03\pm 0.02$ \\
 \HdA~(\AA)     & $c_2$ & $-4.48\pm 0.06$ & $-3.83\pm 0.10$ & $-4.70\pm 0.10$ & $-4.79\pm 0.15$ \\
 & & & & & \\
 \HdF~(\AA)     & $c_1$ & $ 0.60\pm 0.01$ & $ 0.53\pm 0.01$ & $ 0.61\pm 0.01$ & $ 0.67\pm 0.01$ \\
 \HdF~(\AA)     & $c_2$ & $-1.78\pm 0.04$ & $-1.46\pm 0.07$ & $-1.88\pm 0.07$ & $-1.87\pm 0.10$ \\
 & & & & & \\
 \HgA~(\AA)     & $c_1$ & $-5.08\pm 0.01$ & $-5.20\pm 0.02$ & $-5.09\pm 0.01$ & $-4.99\pm 0.02$ \\
 \HgA~(\AA)     & $c_2$ & $-4.67\pm 0.06$ & $-3.99\pm 0.10$ & $-4.80\pm 0.10$ & $-5.15\pm 0.15$ \\
 & & & & & \\
 \HgF~(\AA)     & $c_1$ & $-1.11\pm 0.01$ & $-1.18\pm 0.01$ & $-1.11\pm 0.01$ & $-1.05\pm 0.01$ \\
 \HgF~(\AA)     & $c_2$ & $-2.62\pm 0.04$ & $-2.32\pm 0.06$ & $-2.67\pm 0.06$ & $-2.69\pm 0.09$ \\
 & & & & & \\
 \Fe~(\AA)      & $c_1$ & $ 2.59\pm 0.01$ & $ 2.60\pm 0.01$ & $ 2.58\pm 0.01$ & $ 2.59\pm 0.01$ \\
 \Fe~(\AA)      & $c_2$ & $ 0.93\pm 0.08$ & $ 0.84\pm 0.07$ & $ 0.97\pm 0.08$ & $ 1.01\pm 0.11$ \\
 & & & & & \\
 \Mgb~(\AA)     & $c_1$ & $ 4.00\pm 0.01$ & $ 4.05\pm 0.01$ & $ 3.99\pm 0.01$ & $ 3.91\pm 0.01$ \\
 \Mgb~(\AA)     & $c_2$ & $ 3.44\pm 0.07$ & $ 3.31\pm 0.07$ & $ 3.55\pm 0.08$ & $ 3.31\pm 0.11$ \\
 & & & & & \\
 \MgFep~(\AA)   & $c_1$ & $ 3.25\pm 0.01$ & $ 3.28\pm 0.01$ & $ 3.25\pm 0.01$ & $ 3.21\pm 0.01$ \\
 \MgFep~(\AA)   & $c_2$ & $ 1.96\pm 0.05$ & $ 1.85\pm 0.05$ & $ 2.03\pm 0.06$ & $ 1.97\pm 0.08$ \\
 & & & & & \\
 \Ctwo~(\AA)    & $c_1$ & $ 6.28\pm 0.02$ & $ 6.33\pm 0.03$ & $ 6.21\pm 0.02$ & $ 6.23\pm 0.03$ \\
 \Ctwo~(\AA)    & $c_2$ & $ 5.78\pm 0.15$ & $ 5.51\pm 0.15$ & $ 6.24\pm 0.16$ & $ 5.72\pm 0.23$ \\
 & & & & & \\
 \CAone~(\AA)   & $c_1$ & $ 0.99\pm 0.01$ & $ 1.00\pm 0.01$ & $ 0.98\pm 0.01$ & $ 0.99\pm 0.01$ \\
 \CAone~(\AA)   & $c_2$ & $ 0.58\pm 0.06$ & $ 0.52\pm 0.05$ & $ 0.63\pm 0.06$ & $ 0.46\pm 0.08$ \\

\enddata
\tablecomments{
$\mathrm{Index} = c_1 + c_2~(\log_{10} \sigma - 2.2)$, where $\sigma$ is in $\kms$.
}
\label{licktable}
\end{deluxetable}

\newpage
\clearpage
\begin{deluxetable}{cccccc}
\tabletypesize{\scriptsize}
\tablecolumns{5}
\tablecaption{SSP parameter-$\sigma$ relation}
\tablehead{
\colhead{SSP parameters} & \colhead{Coeffecient} & \colhead{All} & \colhead{Rich group} & \colhead{Poor group} & \colhead{Field}}
\startdata
 $\log_{10}$ Age (Gyr) & $s_1$ & $ 0.78\pm 0.01$ & $ 0.81\pm 0.01$ & $ 0.77\pm 0.01$ & $ 0.74\pm 0.01$ \\
 $\log_{10}$ Age (Gyr) & $s_2$ & $ 0.50\pm 0.04$ & $ 0.45\pm 0.05$ & $ 0.50\pm 0.06$ & $ 0.45\pm 0.08$ \\ 
 & & & & & \\
 $[$Fe/H$]$            & $s_1$ & $-0.06\pm 0.01$ & $-0.06\pm 0.01$ & $-0.06\pm 0.01$ & $-0.05\pm 0.01$ \\
 $[$Fe/H$]$            & $s_2$ & $ 0.37\pm 0.03$ & $ 0.36\pm 0.04$ & $ 0.42\pm 0.05$ & $ 0.36\pm 0.06$ \\
 & & & & & \\
 $[$Mg/Fe$]$           & $s_1$ & $ 0.25\pm 0.01$ & $ 0.25\pm 0.01$ & $ 0.24\pm 0.01$ & $ 0.24\pm 0.01$ \\
 $[$Mg/Fe$]$           & $s_2$ & $ 0.32\pm 0.02$ & $ 0.29\pm 0.03$ & $ 0.35\pm 0.04$ & $ 0.27\pm 0.05$ \\
\enddata
\tablecomments{
$\mathrm{SSP~parameters} = s_1 + s_2~(\log_{10} \sigma  - 2.2)$, where $\sigma$ is in $\kms$.
}
\label{ssptable}
\end{deluxetable}

\newpage
\clearpage

\newpage
\clearpage

\newpage
\clearpage

\newpage
\clearpage

\newpage
\clearpage

\newpage
\clearpage

\newpage
\clearpage

\newpage
\clearpage

\newpage
\clearpage

\newpage
\clearpage

\newpage
\clearpage

\newpage
\clearpage

\newpage
\clearpage

\newpage
\clearpage

\newpage
\clearpage

\newpage
\clearpage

\newpage
\clearpage

\newpage
\clearpage

\newpage
\clearpage

\newpage
\clearpage

\newpage
\clearpage

\newpage
\clearpage

\newpage
\clearpage

\end{document}